\documentclass[a4paper, 11pt]{article}

\usepackage{graphicx}
\usepackage{amsmath,amssymb}
\usepackage[margin=30mm]{geometry}
\usepackage{color}
\usepackage{bbm}
\usepackage{here}
\usepackage{comment}

\begin{document}

\begin{titlepage}

\begin{center}

 {\Large \bf Revisiting Atiyah-Hitchin manifold  \\ 
\vskip 0.2cm
in the generalized Legendre transform} 
\vskip 1.0cm
\normalsize

{\bf Masato Arai${}^{a}$\footnote{arai(at)sci.kj.yamagata-u.ac.jp},
Kurando Baba$^b$\footnote{kurando.baba(at)rs.tus.ac.jp}
and Radu A. Iona\c{s}${}^{c}$\footnote{radu.ionas(at)stonybrook.edu}
}

\vskip 0.5cm

{\it $^a$Faculty of Science, Yamagata University, Kojirakawa-machi 1-4-12, Yamagata, 
Yamagata 990-8560, Japan \\
\vskip 0.5cm
$^b$Department of Mathematics, Faculty of Science and Technology, Tokyo University of Science, Noda, Chiba, 278-8510, Japan
\vskip 0.5cm
$^c$ Department of Physics and Astronomy, Stony Brook University, Stony Brook, NY 11794-3800, USA
}
\vskip 2cm

\begin{abstract}
We revisit construction of the Atiyah-Hitchin manifold in the generalized Legendre transform approach.
This is originally studied by Ivanov and Rocek and is subsequently investigated more by Ionas,
 in the latter of which the explicit forms of the K\"ahler potential and the K\"ahler metric are
 calculated.
There is a difference between the former and the latter.
In the generalized Legendre transform approach, a K\"ahler potential is constructed from the contour 
 integration of one function with holomorphic coordinates.
The choice of the contour in the latter is different from the former's one, whose 
 difference may yield a discrepancy in the K\"ahler potential and eventually in the K\"ahler metric.
We show that the former only gives the real K\"ahler potential, which is consistent 
 with its definition, while the latter yields the complex one.
We derive the K\"ahler potential and the metric for the Atiyah-Hitchin manifold 
 in terms of holomorphic coordinates for the contour considered by Ivanov and Ro\v{c}ek for the first time.
\end{abstract}

\begin{flushleft}
{\small Keywords}\\
\small Hyperk\"ahler manifolds; Generalized Legendre transform; K\"ahler potential potential; K\"ahler metric
\end{flushleft}

\end{center}
\end{titlepage}

\tableofcontents

%
%
\section{Introduction}
So far several constructions of hyperk\"ahler metrics have been proposed in extended
 supersymmetric nonlinear sigma models.
One of ways is hyperk\"ahler quotient framework, where the nonlinear sigma models are described by
 dynamical hypermultiplets and an auxiliary gauge supermultiplet (vectormultiplet).
In this framework, the cotangent bundle over ${\mathbb C}P^n$ model 
 \cite{Curtright:1979yz, Alvarez-Gaume:1980xat, Rocek:1980kc} and its generalization such as
 the cotangent bundle over Grassmannian \cite{Lindstrom:1983rt} have been constructed.
Another interesting approach is to use the projective superspace
 formalism \cite{Karlhede:1984vr}, which is an ${\cal N}=2$ off-shell superfield formulation. 
Based on this formulation, constructions of the cotangent bundles over ${\mathbb C}P^1$ 
 \cite{Gates:1998si} and ${\mathbb C}P^n$ \cite{Gates:1999ea} and the other hermitian symmetric spaces 
 \cite{Arai:2006qt, Arai:2007cx, Arai:2012uh, Arai:2014ita} have been worked out.

The other novel construction of hyperk\"ahler metric is based on the generalized Legendre transform 
 approach \cite{Lindstrom:1983rt, Hitchin:1986ea, Karlhede:1986mg, Lindstrom:1987ks}. 
This approach relates the K\"ahler potentials of certain hyperk\"ahler manifolds to a linear space.
It has also connection to the theory of twistor spaces of hyperk\"ahler manifolds \cite{Hitchin:1986ea}.
A merit of the generalized Legendre transform approach is that 
 the complex structure is manifest.
Namely geometrical quantities such as the K\"ahler metric are described by holomorphic coordinates.
In this approach, a K\"ahler potential is constructed from
 the contour integration of one function with holomorphic coordinates (see \eqref{F-func} and \eqref{Kahler1}). 
This integration is called the $F$-function. 
The $F$-function of the several hyperk\"ahler metrics such as
 Eguchi-Hanson family of self-dual instantons \cite{Eguchi:1978gw, Gibbons:1979xm},
 Taub-NUT family of self-dual instantons \cite{Gibbons:1979xm}
 and the metric due to Calabi \cite{calabi} have been constructed \cite{Hitchin:1986ea}.

After the above studies, the Atiyah-Hitchin manifold \cite{Atiyah:1988jp}, which is a metric on the centered 
 moduli space of two Bogomol'nyi-Prasad-Sommerfield $SU(2)$ monopoles, has been also constructed \cite{IR}.
In \cite{IR}, the $F$-function giving the Atiyah-Hitchin manifold is proposed and corresponding 
 K\"ahler $2$-form is derived, which precisely coincides with one of the Atiyah-Hitchin manifold.
On the other hand,  the explicit form of the K\"ahler potential and the metric 
 are not derived there because the calculation becomes complicated and messy.
A different form of the $F$-function of the Atiyah-Hitchin manifold has been also derived through the twistor space 
 in \cite{Cherkis:1998hi}.
In this paper too, the K\"ahler potential and the metric have not been calculated.
The generalization of the Atiyah-Hitchin manifold to the $k$-monopole case \cite{Houghton:1999hr} and the 
 asymptotically locally flat hyperkahler manifold called $D_k$ type \cite{Cherkis:2003wk} have been discussed.
The $k=2$ case for the former and $D_0$ case for the latter corresponds to the Atiyah-Hitchin manifold.
In \cite{Houghton:1999hr}, the K\"ahler potential and the metric have not been derived explicitly while in
 \cite{Cherkis:2003wk} the metric has been evaluated but it is written by integral form.
 The first explicit calculation of the K\"ahler potential and the metric for the 
  Atiyah-Hitchin manifold in the generalized Legendre transform approach has been worked out in \cite{Ionas1, Ionas2}.
In the calculation, the original holomorphic coordinates in the $F$-function 
 are kept and 
 the metric is derived in terms of them.
Deriving the metric where the complex structure is manifestly kept is important,
 because it would be useful to investigate geometrical properties.
However, we stress that several points in 
\cite{Ionas1, Ionas2}
should be reconsidered.
First of all, the contour of the integration in the $F$-function chosen in \cite{Ionas1} is different from 
 one of \cite{IR}
\footnote{The same contour in \cite{Ionas1} is also chosen in other literatures \cite{Bakas:1999wq, Alexandrov:2008ds}
 In \cite{Houghton:1999hr}, it seems that the same contour as in \cite{Ionas1} is chosen. 
The contours in \cite{Cherkis:1998hi, Cherkis:2003wk} cannot be directly compared with one in \cite{Ionas1} 
 since the forms of the $F$-functions are different from one in \cite{Ionas1}.},
 but it can be shown that this choice does not yield a real K\"ahler potential but 
 a complex one.
The former is consistent with the definition of the K\"ahler potential.
In fact, it is possible to show that the choice in \cite{IR} gives a real K\"ahler potential.
Thus, the calculation deriving the K\"ahler potential and the K\"ahler metric in \cite{Ionas1} should be 
 discussed again with the choice of the contour in \cite{IR}.  
In addition, the detailed derivation of the K\"ahler potential and the K\"ahler metric and the proof of the 
 necessary formulas for the calculation should be also included.
In the derivation in \cite{Ionas1, Ionas2}, the elliptic integrals, their differentiation,
 and their related formulas are heavily used, 
 they are partially explained in \cite{Ionas1} and are not sufficient. 
Therefore, they should be explained in a comprehensive way. 
Moreover, some formulas for the derivation and their proof should be incorporated and placed properly.
Indeed, the formulas necessary in \cite{Ionas1} are given in \cite{Ionas2} and the proof of the formulas is not 
 fully provided.
We emphasize that it is difficult to derive the K\"ahler manifold and the metric for the Atiyah-Hitchin
 manifold for the contour in \cite{IR} simply by referring \cite{Ionas1, Ionas2} and that a detailed explicit calculation 
 to derive them is necessary.

In this paper, we restudy the Atiyah-Hitchin manifold in the generalized Legendre transform approach.
We employ the contour of the integration in the $F$-function in \cite{IR} and derive the K\"ahler potential 
 and the K\"ahler metric with holomorphic coordinates for that choice of the contour for the
 first time.
We show that the choice of the contour in \cite{IR} gives a real K\"ahler potential.
We provide all the necessary steps to derive 
the K\"ahler potential and the metric,
starting from the $F$-function  
 (see \eqref{eqn:Fdef}) which defines the Atiyah-Hitchin manifold.
The formulas related to elliptic integrals, elliptic functions and the other formulas are given in the main 
 body and Appendices.
We find that the resultant K\"ahler potential and the metric for the Atiyah-Hitchin manifold (which are 
 given in \eqref{eqn:fnresult_K} and \eqref{eqn:KZZb}-\eqref{eqn:KUUb}) 
 slightly different from ones in \cite{Ionas1}.
The coefficients of them are different.
This stems from the difference of the choice of the contour. 

This paper is organized as follows. 
In Sec. \ref{Sec:GLT}, we briefly review the generalized Legendre transform approach.
In Sec. \ref{Sec:AH}, we give the $F$-function defining the Atiyah-Hitchin manifold 
and perform the integration in the $F$-function by using the theory of Weierstrass elliptic function.
Finally we derive the K\"ahler potential and the K\"ahler metric by means of the elliptic integrals.
Sec. \ref{sec:con} is denoted to conclusion.
In Appendix \ref{sec:Weierstrass}, we summarize about Weierstrass $\wp $-function, $\zeta $-function 
 and $\sigma $-functions.
In Appendix \ref{sec:pr}, the proofs of relations used in the calculation of the K\"ahler potential
 are given.
In Appendix \ref{sec:dfw}, differential formulas in Weierstrass normal form which is necessary to derive
 the K\"ahler metric are explained.
%
%
\section{The generalized Legendre transform}\label{Sec:GLT}
We briefly review of the generalized Legendre transform construction of hyperk\"ahler manifold \cite{Lindstrom:1987ks}.
We start with a polynomial 
\begin{eqnarray}
 \eta^{(2j)}={\bar{z} \over \zeta^j}+{\bar{v} \over \zeta^{j-1}}+{\bar{t} \over \zeta^{j-2}}+\cdots +x
  +(-)^j(\cdots + t \zeta^{j-2}-v \zeta^{j-1}+z\zeta^j)\,, \label{eta2j}
\end{eqnarray}
where $z, t, \cdots, x$ are holomorphic coordinates and $\zeta$ is the coordinate of the Riemann 
 sphere $\mathbb{C}P^1=S^2$.
This polynomial is called an ${\cal O}(2j)$-multiplet.
Eq. (\ref{eta2j}) should obey the reality condition
\begin{eqnarray}
\eta^{(2j)}(-1/\bar{\zeta})=\overline{\eta^{(2j)}(\zeta)}\,. \label{real}
\end{eqnarray}
The K\"ahler potential for a hyperk\"ahler manifold is constructed from a function with $\eta^{(2j)}$:
\begin{eqnarray}
 F=\oint_C {d\zeta \over \zeta}G(\eta^{(2j)})\,, \label{F-func}
\end{eqnarray}
where $G$ is an arbitrary holomorphic (possibly single or multi-valued) function and the contour $C$ is chosen 
 such that the result of the integration is real.
We call \eqref{F-func} the $F$-function.
The $F$-function satisfies the following set of second order differential equations
\begin{eqnarray}
 &&F_{z\bar{z}}=-F_{v\bar{v}}=F_{t\bar{t}}=\cdots =(-)^jF_{xx}\,, \\
 &&F_{z\bar{v}}=-F_{v\bar{t}}=\cdots\,,\\
 &&F_{zt}=F_{vv}\quad \quad {\rm etc}\,,\\
 &&F_{zv}=F_{vz}\quad \quad {\rm etc}\,,
\end{eqnarray}
where 
\begin{eqnarray}
F_{z\bar{z}}\equiv {\partial F^2 \over \partial z \partial {\bar{z}}}\,,\quad {\rm etc.}
\end{eqnarray}
The K\"ahler potential can be constructed from the 
$F$-function
by performing a two dimensional Legendre
 transform with respect to $v$ and $\bar{v}$
 \begin{eqnarray}
  K(z,\bar{z},u,\bar{u})=F(z,\bar{z},v,\bar{v},t,\bar{t},\cdots,x)-uv-\bar{u}\bar{v}\,, \label{Kahler1}
 \end{eqnarray}
 together with the extremizing conditions
 \begin{eqnarray}
  &&{\partial F \over \partial v}=u\,, \label{Fv}\\
  &&{\partial F \over \partial t}=\cdots={\partial F \over \partial x}=0 \label{Ft}\,.
 \end{eqnarray}
These equations tell us that $v, \bar{v}, t, \bar{t},\cdots,x$ are implicit functions of $z,\bar{z},u,\bar{u}$.
Considering that fact,  differentiating (\ref{Fv}) and (\ref{Ft}) with respect to $z$ gives
\begin{eqnarray}
  F_{zb}+{\partial a \over \partial z}F_{ab}=0\,,  \label{Fzb}
\end{eqnarray}
where $a, b$ run over $v,\bar{v},t,\bar{t},\cdots,x$ and summation over repeated indices is assumed.
Eq. (\ref{Fzb}) yields
\begin{eqnarray}
{\partial a \over \partial z}=-F^{ab}F_{bz}\,, \label{az}
\end{eqnarray}
where we have used $F_{zb}=F_{bz}$ and $F^{ab}$ is the inverse matrix of $F_{ab}$.
On the other hand, differentiating (\ref{Ft}) with respect to $u$, we have
\begin{eqnarray}
{\partial a \over \partial u}=F^{av}\,. \label{au}
\end{eqnarray}
Eqs. (\ref{az}) and (\ref{au}) are used to derive the K\"ahler metric in terms of derivatives of $F$ with
 respect to the holomorphic coordinates.
Taking the derivatives of (\ref{Kahler1}) with respect to $z$ and $u$, we obtain
\begin{eqnarray}
 &\displaystyle {\partial K \over \partial z}={\partial F \over \partial z}\,, &\label{Kz} \\
 &\displaystyle {\partial K \over \partial u}=-v\,. &\label{Ku} 
\end{eqnarray}
Further taking the derivatives of (\ref{Kz}) and (\ref{Ku}) 
and using (\ref{az}) and (\ref{au}), 
 we have the K\"ahler metric as
\begin{eqnarray}
&&K_{z\bar{z}}=F_{z\bar{z}}-F_{za}F^{ab}F_{b\bar{z}}\,, \label{eqn:Kzzbar-F}\\
&&K_{z\bar{u}}=F_{za}F^{a\bar{v}}\,,\\
&&K_{u\bar{z}}=F^{va}F_{a\bar{z}}\,, \\
&&K_{u\bar{u}}=-F^{v\bar{v}}\,. \label{eqn:Kuubar-F}
\end{eqnarray}
%
%
%
\section{The Atiyah-Hitchin metric}\label{Sec:AH}
In this section we give 
the $F$-function
for the Atiyah-Hitchin manifold and derive
 the K\"ahler potential and the  K\"ahler metric.
For the Atiyah-Hitchin manifold, the polynomial in (\ref{F-func}) is an $\mathcal{O}(4)$-multiplet.
In derivation of the K\"ahler potential and the  K\"ahler metric, we heavily use 
the elliptic integrals,
the elliptic functions and their relations.
The formulas are written in the main body and the necessary proofs of the relations are given in
Appendices. 
%
%
%
\subsection{The function $F$ for  the Atiyah-Hitchin manifold}\label{sec:function-FAH}
Let us consider an $\mathcal{O}(4)$-multiplet $\eta^{(4)}=\eta^{(4)}(\zeta)$ expressed in
a Majorana normal form:
\begin{align}\label{eqn:O4multiplet}
\eta^{(4)}
=\dfrac{\bar{z}}{\zeta^{2}}+\dfrac{\bar{v}}{\zeta}+x-v\zeta+z\zeta^{2}\,.
\end{align}
It is convenient to rewrite this form in terms of its roots and a scale factor.
Since it obeys the reality condition (\ref{real}),
the four roots of $\eta^{(4)}$ are invariant under the antipodal map $\zeta\mapsto -1/\bar{\zeta}$.
Hence \eqref{eqn:O4multiplet} is expressed as
\begin{equation}\label{eqn:multiplet_rf}
\eta^{(4)}
=\dfrac{\rho}{\zeta^{2}}\dfrac{(\zeta-\alpha)(\bar{\alpha}\zeta+1)}{(1+|\alpha|^{2})}
\dfrac{(\zeta-\beta)(\bar{\beta}\zeta+1)}{(1+|\beta|^{2})}\,.
\end{equation}
The relations between $z,v,x$ and the roots are obtained by comparing \eqref{eqn:O4multiplet} with \eqref{eqn:multiplet_rf}
 as
\begin{align}
z &= \dfrac{\rho\bar{\alpha}\bar{\beta}}{(1+|\alpha|^{2})(1+|\beta|^{2})}\,,\label{eqn:z}\\
v &= \dfrac{-\rho(\bar{\alpha}+\bar{\beta}-|\alpha|^{2}\bar{\beta}-\bar{\alpha}|\beta|^{2})}{(1+|\alpha|^{2})(1+|\beta|^{2})}\,,\label{eqn:v}\\
x &= \dfrac{\rho(-\bar{\alpha}\beta-\alpha\bar{\beta}+(1-|\alpha|^{2})(1-|\beta|^{2}))}{(1+|\alpha|^{2})(1+|\beta|^{2})}\,.\label{eqn:x}
\end{align}
Since $x$ is real, so is the scale factor $\rho$ by \eqref{eqn:x}.
Without loss of generalities we may assume that the scale factor $\rho$ is positive.

\begin{figure}[H]
  \centering
  \includegraphics[width=7cm]{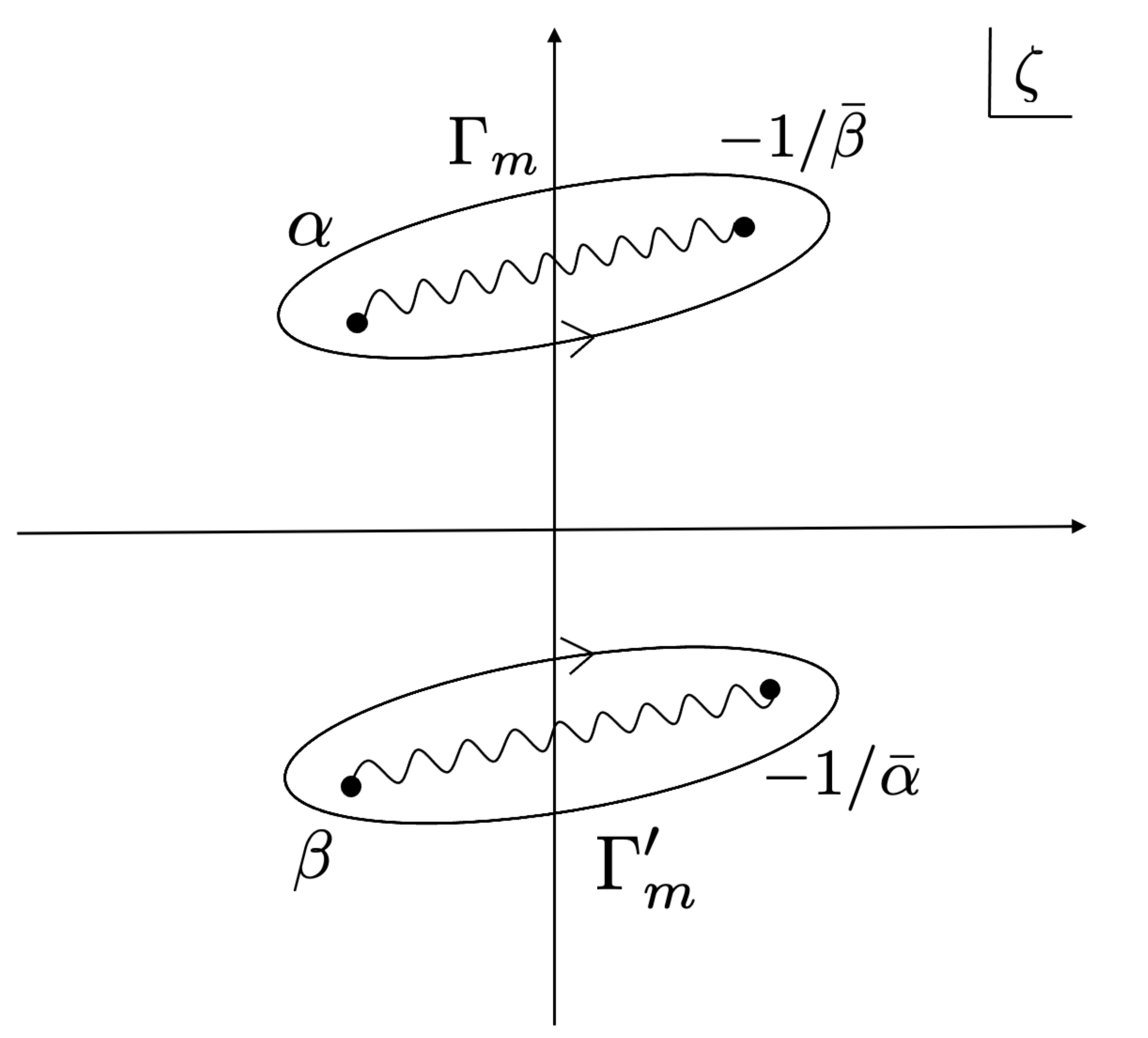}
  \caption{Contour $\Gamma$}\label{contour1}
\end{figure}

Following to \cite{IR} and \cite{Ionas1}, the $F$-function, $F=F(z,\bar{z},v,\bar{v},x)$, of the Atiyah-Hitchin manifold
 is given by\footnote{We follow the sign in \cite{IR} while in \cite{Ionas1} the overall sign is different.}
\begin{equation}\label{eqn:Fdef}
F=F_{2}+F_{1}
=-\dfrac{1}{2\pi i h}\oint_{\Gamma_{0}}\dfrac{d\zeta}{\zeta}\eta^{(4)}
+\oint_{\Gamma}\dfrac{d\zeta}{\zeta}\sqrt{\eta^{(4)}}\,,
\end{equation}
where $h$ is a constant coupling scale, $\Gamma_{0}$ is an integration contour
 encircling the origin of $\zeta$-plane in a counterclockwise direction, and $\Gamma=\Gamma_{m}\cup\Gamma_{m}'$ 
 is a double contour that winds once around two branch-cuts between $\alpha$ and $-1/\bar{\beta}$,
and $\beta$ and $-1/\bar{\alpha}$ (see Fig.~\ref{contour1}).
As we will see later, the contour $\Gamma_{m}$ corresponds to a meridian on a torus associated with $\eta^{(4)}$ 
 (see \eqref{eqn:torus} for the definition of the torus).
Then, we find that $\Gamma_{m}'$ is homologous to $\Gamma_{m}$ on the torus.
Note that the choice of the double contour is firstly proposed in \cite{IR} while only $\Gamma_m$ is chosen as $\Gamma$,
 which we call a single contour, in \cite{Ionas1} as well as the other literatures such as \cite{Bakas:1999wq, Alexandrov:2008ds}.
In the case of the single contour, it can be shown that the function $F_1$ is not real and eventually the K\"ahler potential 
 is not real. 
However, the double contour makes  $F_1$ and the K\"ahler potential real-valued. 
We shall give the proof for completeness.
The function $F_{1}=F_{1}(z,\bar{z},v,\bar{v},x)$ satisfies
\begin{equation}
F_{1}(\lambda z,\lambda \bar{z},\lambda v,\lambda \bar{v},\lambda x)
=\lambda^{1/2}F_{1}(z,\bar{z},v,\bar{v},x)\,,
\end{equation}
for any $\lambda\in\mathbb{R}$, so that, by applying Euler's homogeneous function theorem to $F_{1}$,
we obtain
\begin{equation}\label{eqn:F2Ehft}
F_{1}=
2z\dfrac{\partial F_{1}}{\partial z}
+2\bar{z}\dfrac{\partial F_{1}}{\partial \bar{z}}
+2v\dfrac{\partial F_{1}}{\partial v}
+2\bar{v}\dfrac{\partial F_{1}}{\partial \bar{v}}
+2x\dfrac{\partial F_{1}}{\partial x}\,.
\end{equation}
If we put, for $n\in\mathbb{Z}$,
\begin{equation}\label{eqn:defIn}
\mathcal{I}_{n}=\oint_{\Gamma}\zeta^{n}\dfrac{d\zeta}{2\zeta\sqrt{\eta^{(4)}}}\,,
\end{equation}
then the partial derivatives in \eqref{eqn:F2Ehft} are rewritten as follows:
\begin{equation}\label{eqn:delF1}
\dfrac{\partial F_{1}}{\partial z}=\mathcal{I}_{2}\,,\quad
\dfrac{\partial F_{1}}{\partial \bar{z}}=\mathcal{I}_{-2}\,,\quad
\dfrac{\partial F_{1}}{\partial v}=-\mathcal{I}_{1}\,,\quad
\dfrac{\partial F_{1}}{\partial \bar{v}}=\mathcal{I}_{-1}\,,\quad
\dfrac{\partial F_{1}}{\partial x}=\mathcal{I}_{0}\,.
\end{equation}
Thanks to the choice of $\Gamma$ we can prove the following relation
\footnote{A similar relation for the single contour is given in \cite{Ionas1}, where there is an additional
 term in the right-hand side in (\ref{eqn:InbarIn}). In our case, such a term does not exist due to the choice
 of the double contour.}
:
\begin{equation}\label{eqn:InbarIn}
\mathcal{I}_{-n}
=(-1)^{n}\overline{\mathcal{I}_{n}}\,.
\end{equation}
Its proof is given in Appendix \ref{sec:InbarIn}.
It follows from \eqref{eqn:InbarIn} that $\mathcal{I}_{0}$ is real-valued,
and that \eqref{eqn:F2Ehft} is rewritten as
\begin{equation}
F_{1}
=2x\mathcal{I}_{0}-2(v\mathcal{I}_{1}+\overline{v\mathcal{I}_{1}})+2(z\mathcal{I}_{2}+\overline{z\mathcal{I}_{2}})\,.
\end{equation}
Therefore we have shown that $F_{1}$ is real-valued.
As will be shown in \eqref{eqn:f2_xh}, since $F_2$ is also real-valued,
from \eqref{Kahler1}, the K\"ahler potential is real-valued.

In order to derive the K\"ahler potential from (\ref{eqn:Fdef}) by the generalized Legendre transformation, we need to
 perform the integrals in (\ref{eqn:Fdef}). Explicitly they are $F_2$, and ${\cal I}_n(n=0,1,2)$ in $F_1$.
$F_2$ can be evaluated by means of a straightforward application of Cauchy's integral formula.
Then, we get
\begin{equation}\label{eqn:f2_xh}
F_{2}=-\dfrac{x}{h}\,.
\end{equation}

\begin{figure}[H]
  \centering
  \includegraphics[width=13cm]{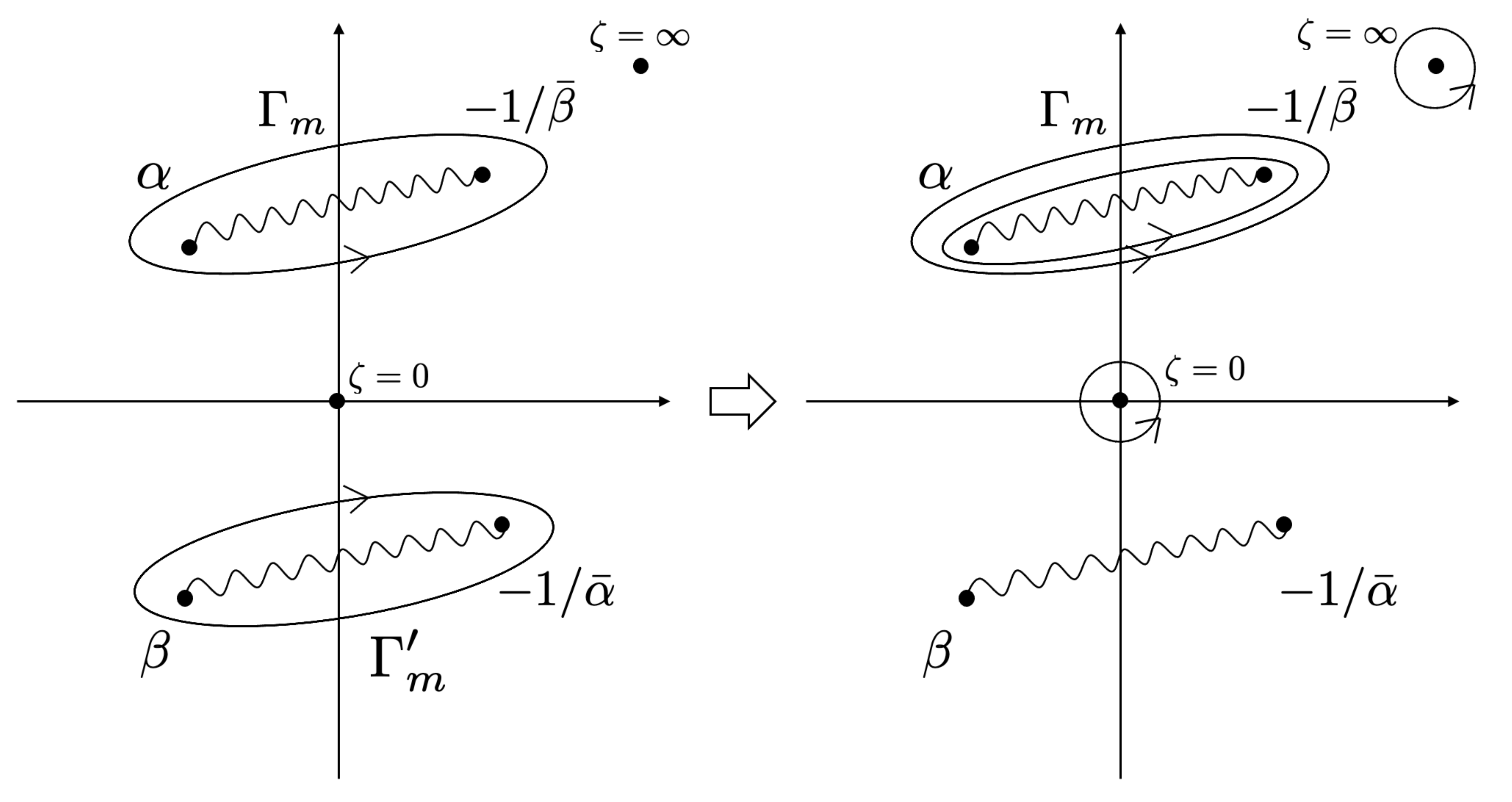}
  \caption{Deformation of $\Gamma_{m}'$ to $\Gamma_{m}$}\label{fig:def}
\end{figure}

The evaluation of ${\cal I}_n$ in $F_1$ needs several steps.
First of all, let us rewrite ${\cal I}_n$ in terms of only the single contour $\Gamma_m$.
When deforming $\Gamma_{m}'$ to $\Gamma_{m}$, we need to pick up the residues of
 the integrand of $\mathcal{I}_{n}$ 
(see Fig.~\ref{fig:def}).
\noindent
Since this integrand has two simple poles (except $n=0$ case, see \eqref{eqn:I12res}); 
 one at $\zeta=0$ and the other at $\zeta=\infty$, we obtain
\begin{equation}\label{eqn:Indtos}
\mathcal{I}_{n}
=\left(\oint_{\Gamma_{m}}+\oint_{\Gamma_{m}'}\right)\zeta^{n}\dfrac{d\zeta}{2\zeta\sqrt{\eta^{(4)}}}
=2\oint_{\Gamma_{m}}\zeta^{n}\dfrac{d\zeta}{2\zeta\sqrt{\eta^{(4)}}}
+2\pi i\left\{R(0,n)+R(\infty,n)\right\}\,,
\end{equation}
where $R(\zeta,n)$ denotes the residue for the integrand of $\mathcal{I}_{n}$ at $\zeta\in\{0,\infty\}$.
In order to evaluate the right-hand side of \eqref{eqn:Indtos}, we shall obtain the residues $R(\zeta,n)$. 
The integrand of $\mathcal{I}_{n}$ can be expanded around $\zeta=0$ as follows:
\begin{equation}
\zeta^{n}\dfrac{1}{2\zeta\sqrt{\eta^{(4)}}}
=\dfrac{\zeta^{n}}{2\sqrt{\bar{z}}}\left\{
1-\dfrac{\bar{v}}{2\bar{z}}\zeta
+\left(\dfrac{3\bar{v}}{8\bar{z}^{2}}-\dfrac{x}{2\bar{z}}\right)\zeta^{2}
+\dotsb
\right\}\,,
\end{equation}
from which we get
\begin{equation}\label{eqn:R0,n}
R(0,n)=0\,.
\end{equation}
The residue at $\zeta=\infty$ can be computed by changing the variable $\zeta=1/\zeta'$.
If we denote by $\Gamma'$ which is the image of $\Gamma$ mapped by $\zeta=1/\zeta'$
 on the $\zeta'$-plane, then $\mathcal{I}_{n}$ can be rewritten as
\begin{equation}\label{eqn:Inzeta'}
\mathcal{I}_{n}
=\oint_{\Gamma'}
(-\zeta'^{-n})
\dfrac{d\zeta'}{2\zeta'\sqrt{\eta^{(4)}(\zeta')}}\,.
\end{equation}
Since the expansion of its integrand around $\zeta'=0$ is given by
\begin{equation}
(-\zeta'^{-n})
\dfrac{1}{2\zeta'\sqrt{\eta^{(4)}(\zeta')}}
=\dfrac{-\zeta'^{-n}}{2\sqrt{z}}\left\{
1+\dfrac{v}{2z}\zeta'+\left(
\dfrac{3v^{2}}{8z^{2}}-\dfrac{x}{2z}
\right)\zeta'^{2}+\dotsb
\right\}\,,
\end{equation}
we obtain
\begin{equation}\label{eqn:Rinftyn}
R(\infty,n)=\begin{cases}
0 & (n=0)\,, \\
-\dfrac{1}{2\sqrt{z}} & (n=1)\,,\\
-\dfrac{1}{2\sqrt{z}}\cdot\dfrac{v}{2z} & (n=2)\,.\rule{0pt}{4ex}
\end{cases}
\end{equation}
%
Substituting \eqref{eqn:R0,n} and \eqref{eqn:Rinftyn} into \eqref{eqn:Indtos}, we obtain
\begin{equation}\label{eqn:I12res}
\mathcal{I}_{n}
=
\begin{cases}
2\displaystyle\oint_{\Gamma_{m}}\dfrac{d\zeta}{2\zeta\sqrt{\eta^{(4)}}} & (n=0)\,, \\
2\displaystyle\oint_{\Gamma_{m}}\zeta\dfrac{d\zeta}{2\zeta\sqrt{\eta^{(4)}}}
+2\pi i\left(-\dfrac{1}{2\sqrt{z}}\right) & (n=1)\,,\\
2\displaystyle\oint_{\Gamma_{m}}\zeta^{2}\dfrac{d\zeta}{2\zeta\sqrt{\eta^{(4)}}}
+2\pi i\left(-\dfrac{1}{2\sqrt{z}}\cdot\dfrac{v}{2z}\right) & (n=2)\,.\rule{0pt}{4ex}
\end{cases}
\end{equation}
From the above argument, $F_{1}$ is rewritten as \footnote{In \cite{Ionas1}, the residue contribution
 does not exist since the single contour for the function $F$ is considered.}
\begin{align}
F_{1}
&=2x\mathcal{I}_{0}-2\left(v\mathcal{I}_{1}-z\mathcal{I}_{2}+\mathrm{c.c.}\right) \nonumber \\
&= 4\left\{
x\mathcal{I}_{0}(\Gamma_{m})-\left(
v\mathcal{I}_{1}(\Gamma_{m})
-z\mathcal{I}_{2}(\Gamma_{m})-\dfrac{\pi i}{4}\cdot\dfrac{v}{\sqrt{z}}+\mathrm{c.c.}
\right)\right\}\,,\label{eqn:F1_exp}
\end{align}
where we put
\begin{equation}
\mathcal{I}_{n}(\Gamma_{m})
=\oint_{\Gamma_{m}}\zeta^{n}\dfrac{d\zeta}{2\zeta\sqrt{\eta^{(4)}}}\quad
(n=0, 1,2)\,.
\end{equation}
In the following subsections, we will evaluate ${\cal I}_{n}(\Gamma_m)$.

%
%
\subsection{Calculation of ${\cal I}_0(\Gamma_m)$}\label{sec:I0}
\begin{figure}[H]
  \centering
  \includegraphics[width=8cm]{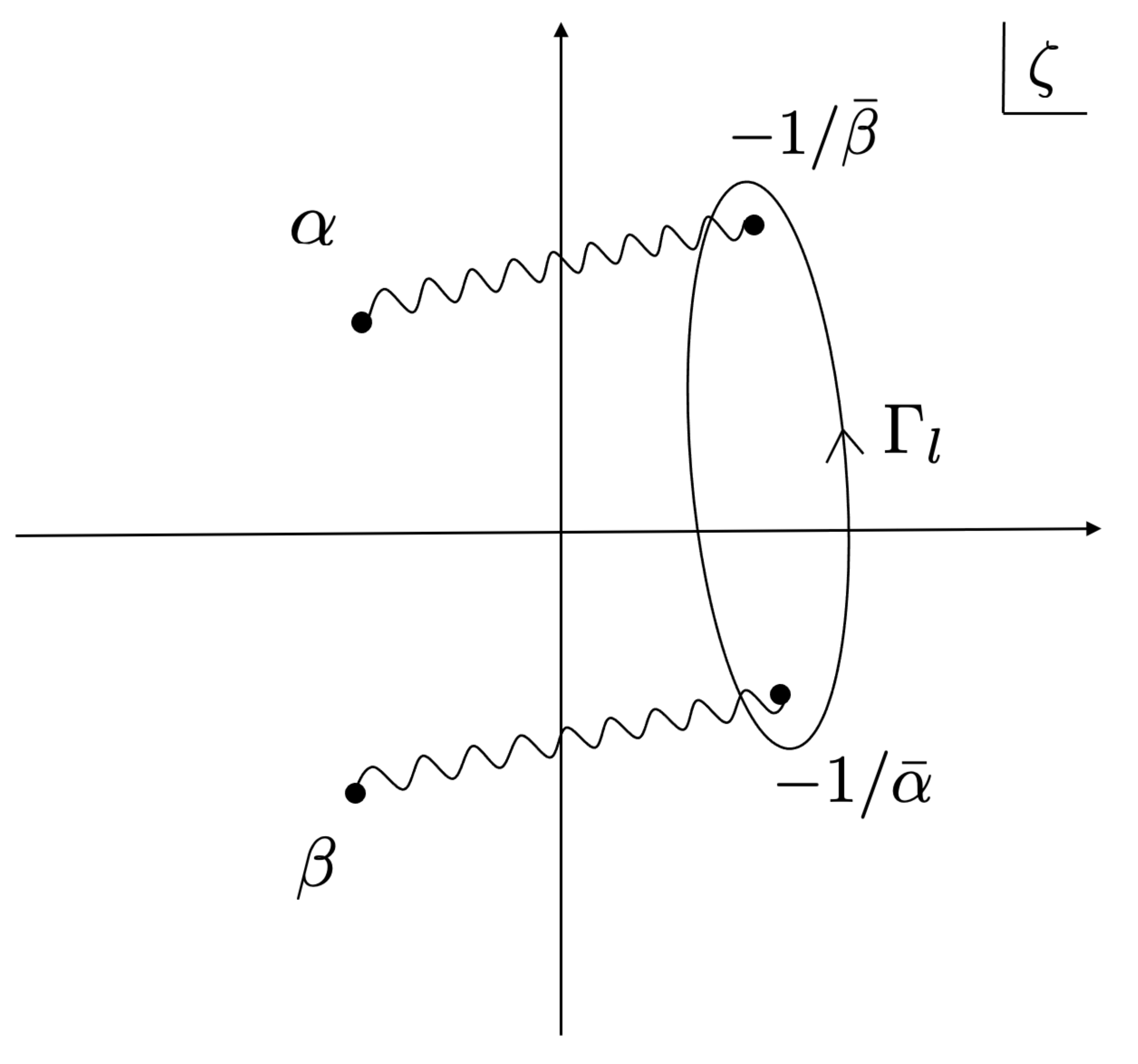}
  \caption{Contour $\Gamma_{l}$}\label{fig:contour3}
\end{figure}

We first consider a $(\zeta,\eta)$-plane curve defined by
\begin{equation}\label{eqn:torus}
C:\quad \eta^{2}=4\zeta^{2}\eta^{(4)}(\zeta)\,.
\end{equation}
It is verified that the projectivization of $C$ is isomorphic to a torus,
 so that this elliptic curve defines a globally defined holomorphic $1$-from $\varpi$
 as follows:
\begin{equation}
\varpi=\dfrac{d\zeta}{\eta}=\dfrac{d\zeta}{2\zeta\sqrt{\eta^{(4)}}}\,. \label{one-form}
\end{equation}
This is the integrand in ${\cal I}_0(\Gamma_m)$. 
Such a form is uniquely determined up to a constant multiple.
We call this form the abel form of $C$.
The two periods of $\varpi$ are described by its integral over canonical cycles.
Namely, one period is equal to
\begin{equation}\label{eqn:omega}
2\omega=\oint_{\Gamma_{m}}\varpi\,,
\end{equation}
and the other is
\begin{equation}\label{eqn:omega'}
2\omega'=\oint_{\Gamma_{l}}\varpi\,,
\end{equation}
where
$\Gamma_{l}$ is a contour that
winds once around the roots $-1/\bar{\beta}$ and $-1/\bar{\alpha}$
(see Fig.~\ref{fig:contour3}), which corresponds to a longitude on the torus.

\noindent
It can be shown that the periods are calculated as
\begin{equation}\label{eqn:omegaKomega'K'}
\omega=\dfrac{1}{\sqrt{\rho}}K(k)\,,
\quad
\omega'=\dfrac{i}{\sqrt{\rho}}K(k')\,,
\end{equation}
where $K(k)$ denotes the complete elliptic integral of the first kind with modulus
\begin{equation}\label{eqn:k_def}
k=\dfrac{|1+\bar{\alpha}\beta|}{\sqrt{(1+|\alpha|^{2})(1+|\beta|^{2})}}\,,
\end{equation}
and $k'=\sqrt{1-k^{2}}$ denotes the complementary modulus.
We give the details of the derivation of (\ref{eqn:omegaKomega'K'}).
We make use of the following birational transformation:
\begin{equation}\label{eqn:bitrans}
\begin{cases}
\nu=[\zeta,-1/\bar{\alpha},\alpha,\beta]
=\dfrac{(\zeta-\alpha)(1+\bar{\alpha}\beta)}{(\zeta-\beta)(1+|\alpha|^{2})}\,,\\
\mu=\eta\dfrac{\partial\nu}{\partial\zeta}\,.
\end{cases}
\end{equation}
Then the curve $C$ in \eqref{eqn:torus}
is expressed as a Riemann normal form (cf.~\cite[p.~64]{Ak}):
\begin{equation}\label{eqn:CLow}
\mu^{2}=4\rho\nu(\nu-1)(\nu-k^{2})\,,
\end{equation}
and the four roots $\alpha$, $-1/\bar{\beta}$, $-1/\bar{\alpha}$,
$\beta$ correspond to $0$, $k^{2}$, $1$, $\infty$, respectively.
In addition, the abel form $\varpi$ is expressed by
\begin{equation}
\varpi
=\dfrac{d\nu}{\mu}
=\dfrac{d\nu}{2\sqrt{\rho\nu(\nu-1)(\nu-k^{2})}}\,.
\end{equation}
Hence we obtain
\begin{align}
\omega
&=\int_{0}^{k^{2}}\dfrac{d\nu}{2\sqrt{\rho\nu(\nu-1)(\nu-k^{2})}}
=\dfrac{1}{\sqrt{\rho}}\int_{0}^{1}\dfrac{dt}{\sqrt{(1-t^{2})(1-k^{2}t^{2})}}
=\dfrac{1}{\sqrt{\rho}}K(k)\,.
\end{align}
Here, we have changed the variable from $\nu$ to t by $\nu=k^{2}t^{2}$
in the second equality.
A similar calculation shows the second equality in \eqref{eqn:omegaKomega'K'}.
Indeed, we observe
\begin{align}
\omega' 
&=\int_{1}^{k^{2}}\dfrac{d\nu}{2\sqrt{\rho\nu(\nu-1)(\nu-k^{2})}}
=\dfrac{i}{\sqrt{\rho}}\int_{0}^{1}\dfrac{dx}{\sqrt{(k'^{2}x^{2}+k^{2})(1-x^{2})}}\,.
\end{align}
Here, in the second equality, we have changed
the variable from $\nu$ to $x$ by
\begin{equation}
x=\sqrt{\dfrac{\nu-k^{2}}{k'^{2}}}\,.
\end{equation}
Using
\begin{equation}
\dfrac{1}{\sqrt{(k'^{2}x^{2}+k^{2})(1-x^{2})}}
=\dfrac{1}{\sqrt{k^{2}}}\cdot
\dfrac{1}{\sqrt{(1+\kappa^{2}x^{2})(1-x^{2})}}\,,\quad
\kappa:=\dfrac{k'}{k}\,,
\end{equation}
and making substitution
\begin{equation}
x=\dfrac{t}{\sqrt{1+\kappa^{2}(1-t^{2})}}\,,
\end{equation}
we have
\begin{equation}
\omega'
=\dfrac{i}{\sqrt{\rho}}\int_{0}^{1}
\dfrac{dt}{\sqrt{(1-t^{2})(1-(\frac{\kappa}{\sqrt{1+\kappa^{2}}})^{2}t^{2})}}
=\dfrac{i}{\sqrt{\rho}}K(k')\,.
\end{equation}
Thus, we have verified \eqref{eqn:omegaKomega'K'}.

The period (\ref{eqn:omega}) is actually ${\cal I}_0(\Gamma_m)$, so we have
\begin{eqnarray}
 {\cal I}_0(\Gamma_m)=\oint_{\Gamma_m}\varpi=2\omega\,. \label{eq:I1}
\end{eqnarray}
%

%
%
\subsection{Calculation of ${\cal I}_1(\Gamma_m)$}
In order to evaluate ${\cal I}_1(\Gamma_m)$ and ${\cal I}_2(\Gamma_m)$, 
 we heavily use
 the theory of Weierstrass elliptic functions
(see \eqref{eqn:piinfty_zetaomega''} and \eqref{eqn:I2Gm_last} for our results).
Our conventions for the elliptic functions are those of \cite{Ak}.
We start with rewriting  the elliptic curve $C$ in \eqref{eqn:torus}
by means of the following transformation
\begin{equation}\label{eqn:numuXY}
\begin{cases}
\nu=\dfrac{X}{\rho}+\dfrac{1+k^{2}}{3}\,,\\
\mu=\dfrac{Y}{\rho}\,.\rule{0pt}{4ex}\,
\end{cases}
\end{equation}
%
By using this transformation, we see that the three roots $1$, $k^{2}$, $0$ 
 of the curve correspond to
\begin{equation}
e_{1}=-\dfrac{\rho}{3}(k^{2}-2)\,,
\quad
e_{2}=\dfrac{\rho}{3}(2k^{2}-1)\,,
\quad
e_{3}=-\dfrac{\rho}{3}(k^{2}+1)\,,
\end{equation}
respectively.
Then, $0<k^{2}<1$ implies $e_{3}<e_{2}<e_{1}$.
Clearly, we get
\begin{equation}
e_{1}-e_{3}=\rho\,,\quad
\dfrac{e_{2}-e_{3}}{e_{1}-e_{3}}=k^{2}\,.
\end{equation}
Hence, the curve $C$ expressed in \eqref{eqn:CLow} becomes
a Weierstrass normal form:
\begin{equation}\label{eqn:Wcurve}
Y^{2}=4(X-e_{1})(X-e_{2})(X-e_{3})=4X^{3}-g_{2}X-g_{3}\,.
\end{equation}
The relation between roots and coefficients are given by
\begin{equation}\label{eqn:rlt_sc}
e_{1}+e_{2}+e_{3}=0\,,\quad
e_{1}e_{2}+e_{2}e_{3}+e_{3}e_{1}=-\dfrac{g_{2}}{4}\,,\quad
e_{1}e_{2}e_{3}=\dfrac{g_{3}}{4}\,.
\end{equation}
This obeys
\begin{equation}\label{eqn:g2g3_def}
g_{2}=\dfrac{4}{3}\rho^{2}(1-k^{2}+k^{4})\,,\quad
g_{3}=\dfrac{4}{27}\rho^{3}(k^{2}-2)(2k^{2}-1)(k^{2}+1)\,.
\end{equation}
Furthermore, the discriminant $\Delta$ of $C$ is not equal to zero because
\begin{equation}\label{eqn:Deltakrho_apndx}
\Delta=
g_{2}^{3}-27g_{3}^{2}
=16\rho^{6}k^{4}k'^{4}\neq 0\,.
\end{equation}
The abel form $\varpi$ is now expressed by
\begin{equation}
\varpi
=\dfrac{dX}{Y}
=\dfrac{dX}{\sqrt{4X^{3}-g_{2}X-g_{3}}}\,.
\end{equation}
Here, we give a summary of the relations among
Majonara normal form, Riemann normal form and Weierstrass normal form
as in Table \ref{table:MLW}.

\begin{table}[H]
\centering
\caption{Majonara normal form,
Riemann normal form
and Weierstrass normal form}\label{table:MLW}
\begin{tabular}{c|c|c|c}
Normal form & Elliptic curve & Roots & Abel form \\
\hline
Majorana & $\eta^{2}=\zeta^{2}\eta^{(4)}(\zeta)$  &
$\alpha$, $-\dfrac{1}{\bar{\beta}}$, $-\dfrac{1}{\bar{\alpha}}$, $\beta$ & $\dfrac{d\zeta}{\eta}$\rule{0pt}{4ex}\\
Riemann & $\mu^{2}=\rho\nu(\nu-1)(\nu-k^{2})$ & $0$, $k^{2}$, $1$, $\infty$ & $\dfrac{d\nu}{\mu}$ \rule{0pt}{4ex}\\
Weierstrass & $Y^{2}=4X^{3}-g_{2}X-g_{3}$ & $e_{3}$, $e_{2}$, $e_{1}$, $\infty$ & $\dfrac{dX}{Y}$\rule{0pt}{4ex}
\end{tabular}
\end{table}

We shall rewrite ${\cal I}_n(\Gamma_m)$ by using the Weierstrass normal form. 
Combined \eqref{eqn:bitrans} and \eqref{eqn:numuXY}, we have a single transformation
\begin{equation}\label{eqn:zeta_X}
\dfrac{(\zeta-\alpha)(1+\bar{\alpha}\beta)}{(\zeta-\beta)(1+|\alpha|^{2})}=\dfrac{X-e_{3}}{e_{1}-e_{3}}\,.
\end{equation}
We denote by $X_{\zeta}$ the image of $\zeta$ through this birational map.
Under this convention we have
\begin{align}
X_{0}
&= e_{3}+(e_{1}-e_{3})\dfrac{\alpha}{\beta}\dfrac{1+\bar{\alpha}\beta}{1+|\alpha|^{2}}
= e_{3}+\rho\cdot\dfrac{\alpha}{\beta}\dfrac{1+\bar{\alpha}\beta}{1+|\alpha|^{2}}\,,\label{eqn:X0_def}\\
X_{\infty}
&= e_{3}+\rho\cdot\dfrac{1+\bar{\alpha}\beta}{1+|\alpha|^{2}}\,.
\label{eqn:Xinfty_def}
\end{align}
Then, \eqref{eqn:zeta_X} is rewritten as
\begin{equation}\label{eqn:zeta2X}
\zeta=\beta\dfrac{X-X_{0}}{X-X_{\infty}}\,.
\end{equation}
The contours $\Gamma_{m}$ and $\Gamma_{l}$ on $\zeta$-plane
 are mapped to ones on $X$-plane via \eqref{eqn:zeta2X},
 which we write the same symbols, namely, $\Gamma_{m}$ (resp.~$\Gamma_{l}$)
 winds once around the branch-cut between the roots $e_{3}$ and $e_{2}$
 (resp.~the roots $e_{2}$ and $e_{1}$) on $X$-plane.
Therefore,
$\mathcal{I}_{n}(\Gamma_{m})$ has the following expression:
\begin{equation}\label{eqn:InGm}
\mathcal{I}_{n}(\Gamma_{m})
=\oint_{\Gamma_{m}}\left(\beta\dfrac{X-X_{0}}{X-X_{\infty}}\right)^{n}\dfrac{dX}{Y}\,.
\end{equation}
Clearly, \eqref{eq:I1} is rewritten as
\begin{equation}\label{eqn:omegaomega'_XY}
{\cal I}_0(\Gamma_m)=\oint_{\Gamma_{m}}\dfrac{dX}{Y}=2\omega\,.
\end{equation}
In the following, we set 
\begin{eqnarray}\label{eqn:omegadef}
\omega_{1}=\omega\,,\quad \omega_{2}=\omega+\omega'\,,\quad \omega_{3}=\omega'\,.
\end{eqnarray}

For the evaluation of \eqref{eqn:InGm}, we need to use Weierstrass 
 $\wp$-function, $\zeta$-function and $\sigma$-function.
For reader's convenience, we briefly review the basics of them in Appendix \ref{sec:Weierstrass}.
To evaluate \eqref{eqn:InGm}, we first define $u_{\zeta}\in\mathbb{C}/\Lambda$
 for $\zeta\in\mathbb{C}\cup\{\infty\}$, by the following equation
\begin{equation}
X_{\zeta}=\wp(u_{\zeta})\,,
\end{equation}
and we set $Y_{\zeta}=\wp'(u_{\zeta})$.
With the use of $X_\zeta$ and $Y_\zeta$, it can be shown that the following equalities hold:
\begin{equation}\label{eqn:YinfXzeroXinf}
\dfrac{Y_{\infty}}{X_{0}-X_{\infty}}
=2\beta\sqrt{z}\,,\quad
\dfrac{Y_{0}}{X_{\infty}-X_{0}}
=\dfrac{2\sqrt{\bar{z}}}{\beta}\,,
\end{equation}
which are frequently used below.
Their proofs are given in Appendix \ref{sec:Xinfty-X0Yinfty}.

It is convenient to divide $u_{\zeta}$ into the real part
 and the imaginary part with respect to
 the antiholomorphic involution $\zeta\mapsto -1/\bar{\zeta}$
 on $\mathbb{C}\cup\{\infty\}$, that is,
\begin{equation}
u_{\zeta}^{\pm}
=u_{\zeta}\pm u_{-1/\bar{\zeta}}\,,
\end{equation}
so that we have 
\begin{equation}\label{eqn:uinfty_pm}
u_{\infty}^{\pm}=u_{\infty}\pm u_{0}\,.
\end{equation}
We write $(x_{\pm},y_{\pm})$ as the $(X,Y)$-coordinates of the point corresponding to 
$u_{\infty}^{\pm}$ via the abel map $\psi$, \eqref{eqn:abel}. 
Thanks to \eqref{eqn:YinfXzeroXinf}, we can prove that the following relation holds:
\begin{equation}\label{eqn:xpm_escpt}
x_{\pm}
= \dfrac{x\pm 6|z|}{3}\,.
\end{equation}
Its proof is given in Appendix \ref{sec:xpm_escpt}.
The coordinate $y_{\pm}$ is calculated by substituting
$(u,v,w)=(u_{\infty},\pm u_{0},-u_{\infty}^{\pm})$ into \eqref{eqn:wpupw}.
Then we have
\begin{equation}
\det\left(
\begin{array}{ccc}
1 & X_{\infty} & Y_{\infty}\\
1 & X_{0} & \pm Y_{0}\\
1 & x_{\pm} & -y_{\pm}\\
\end{array}
\right)=0\,,
\end{equation}
that is,
\begin{equation}\label{eqn:y_pm_des}
y_{\pm}=
(x_{\pm}-X_{0})\dfrac{Y_{\infty}}{X_{0}-X_{\infty}}
\pm(x_{\pm}-X_{\infty})\dfrac{Y_{0}}{X_{\infty}-X_{0}}\,.
\end{equation}
By using \eqref{eqn:z}, \eqref{eqn:v}, \eqref{eqn:x}, \eqref{eqn:X0_def}, \eqref{eqn:Xinfty_def}, \eqref{eqn:YinfXzeroXinf} and \eqref{eqn:xpm_escpt},
 it can be shown that \eqref{eqn:y_pm_des} is rewritten as
\begin{equation}\label{eqn:ypm}
\quad
y_{+} = iv_{+}(x_{+}-x_{-}),
\quad
y_{-} =v_{-}(x_{-}-x_{+}), 
\end{equation}
where we put
\begin{equation}\label{eqn:vpvm_def}
v_{+}=\mathrm{Im}\dfrac{v}{\sqrt{z}}\,,\quad
v_{-}=\mathrm{Re}\dfrac{v}{\sqrt{z}}\,.
\end{equation}

Now let us evaluate ${\cal I}_1(\Gamma_m)$.
First, we observe
\begin{align}
\mathcal{I}_{1}(\Gamma_{m})
=\beta\left\{
\oint_{\Gamma_{m}}\dfrac{dX}{Y}
+\dfrac{X_{\infty}-X_{0}}{Y_{\infty}}\oint_{\Gamma_{m}}
\dfrac{Y_{\infty}}{X-X_{\infty}}\dfrac{dX}{Y}
\right\}\,. \label{eqn:InGm2}
\end{align}
The first term in \eqref{eqn:InGm2} can be easily calculated by using \eqref{eqn:omegaomega'_XY}.
In the second term, we need to evaluate the integral
\begin{equation}\label{eqn:piGamma}
\pi(X_{\zeta})
\equiv -
\oint_{\Gamma_{m}}\dfrac{Y_{\zeta}}{X-X_{\zeta}}\dfrac{dX}{Y}\,,
\quad
\zeta\in\mathbb{C}\cup\{\infty\}\,,
\end{equation}
with $\zeta=\infty$.
We calculate it for arbitrary $\zeta$.
Using the abel map \eqref{eqn:abel} and the following formula (cf.~\cite[p.~41]{Ak}):
\begin{equation}\label{eqn:add}
\dfrac{\wp'(v)}{\wp(u)-\wp(v)}
=-\zeta(u+v)+\zeta(u-v)+2\zeta(v)\,,
\end{equation}
we obtain
\begin{align}
\pi(X_{\zeta})
&=-2\int_{e_{3}}^{e_{2}}\dfrac{Y_{\zeta}}{X-X_{\zeta}}\dfrac{dX}{Y} \nonumber \\
&=-2\int_{\omega_{3}}^{\omega_{2}}\dfrac{\wp'(u_{\zeta})}{\wp(u)-\wp(u_{\zeta})}du \nonumber \\
&=-2\int_{\omega_{3}}^{\omega_{2}}\left\{
-\zeta(u+u_{\zeta})+\zeta(u-u_{\zeta})+2\zeta(u_{\zeta})
\right\}du \nonumber \\
&=-2\int_{\omega_{3}}^{\omega_{2}}\left\{
-\zeta(u+u_{\zeta})+\zeta(u-u_{\zeta})\right\}du
-4\omega_{1}\zeta(u_{\zeta})\,.
\end{align}
Here, in the last equality
we have used $\omega_{1}=\omega_{2}-\omega_{3}$.
By using the definition of $\sigma$-function given in \eqref{eqn:sigma}, we have
\begin{equation}\label{eqn:piXinfty_sig}
\pi(X_{\zeta})
=2\log\dfrac{\sigma(\omega_{2}+u_{\zeta})}{\sigma(\omega_{2}-u_{\zeta})}-2\log\dfrac{\sigma(\omega_{3}+u_{\zeta})}{\sigma(\omega_{3}-u_{\zeta})}
-4\omega_{1}\zeta(u_{\zeta})\,.
\end{equation}
The first and second terms are evaluated by means of the monodromy property of $\sigma$-function
 for $j=2,3$, \eqref{eqn:monod_j1,3}
%
\begin{align}
\sigma(\omega_{j}+u_{\zeta})
=e^{2\eta_{j}u_{\zeta}}\sigma(\omega_{j}-u_{\zeta})\,,
\end{align}
so that the following relation holds:
\begin{equation}\label{eqn:logmonod}
2\log\dfrac{\sigma(\omega_{j}+u_{\zeta})}{\sigma(\omega_{j}-u_{\zeta})}
=
4\eta_{j}u_{\zeta}\quad
(\mathrm{mod}~2\pi i\mathbb{Z})\,,
\end{equation}
where $\eta_j$ is the quasi-half period defined in \eqref{eqn:eta_dperiod} and \eqref{eqn:eta2}.
Substituting \eqref{eqn:logmonod} into \eqref{eqn:piXinfty_sig},
 we obtain
\begin{align}
\pi(X_{\zeta})
= 4\det\left(
\begin{array}{cc}
u_{\zeta} & \omega_{1} \\
\zeta(u_{\zeta}) & \zeta(\omega_{1})
\end{array}
\right)\quad
(\mathrm{mod}~2\pi i\mathbb{Z})
\,,\label{eqn:piinfty_zetaomega}
\end{align}
where we have used 
\eqref{eqn:eta2}.

Next we evaluate \eqref{eqn:piGamma} in the case for $\zeta=\infty$.
We find that $\zeta(u_{\infty})$ is rewritten as
\begin{equation}\label{eqn:zetauinfty}
\zeta(u_{\infty})
=\dfrac{1}{2}\left\{\zeta(u_{\infty}^{+})+\zeta(u_{\infty}^{-})-\dfrac{Y_{\infty}}{X_{\infty}-X_{0}}\right\}\,.
\end{equation}
Indeed, it follows from the formula \eqref{eqn:add} for $u=u_{\infty},v=u_{0}$ that
\begin{align}
\dfrac{Y_{\infty}}{X_{\infty}-X_{0}}
=\dfrac{\wp'(u_{\infty})}{\wp(u_{\infty})-\wp(u_{0})} 
= \zeta(u_{\infty}^{+})+\zeta(u_{\infty}^{-})-2\zeta(u_{\infty})\,,
\end{align}
where 
we have used \eqref{eqn:uinfty_pm}.
Hence, we get \eqref{eqn:zetauinfty}.
By using \eqref{eqn:zetauinfty}, we have
\begin{align}
&\det\left(
\begin{array}{cc}
u_{\infty} & \omega_{1} \notag\\
\zeta(u_{\infty}) & \zeta(\omega_{1})
\end{array}
\right)\\
&\phantom{hogehoge}=\dfrac{1}{2}\left\{
\det\left(
\begin{array}{cc}
u_{\infty}^{+} & \omega_{1} \\
\zeta(u_{\infty}^{+}) & \zeta(\omega_{1})
\end{array}
\right)+
\det\left(
\begin{array}{cc}
u_{\infty}^{-} & \omega_{1} \\
\zeta(u_{\infty}^{-}) & \zeta(\omega_{1})
\end{array}
\right)+
\omega_{1}\dfrac{Y_{\infty}}{X_{\infty}-X_{0}}\right\}\,,
\end{align}
from which we obtain
\begin{align}
\pi(X_{\infty})
=\dfrac{1}{2}\left\{\pi(x_{+})+\pi(x_{-})\right\}+2\omega_{1}\dfrac{Y_{\infty}}{X_{\infty}-X_{0}}\quad
(\mathrm{mod}~\pi i\mathbb{Z})\,.
\end{align}
Thus,
there exists $a\in\mathbb{Z}$ such that
\begin{equation}\label{eqn:piinfty_zetaomega'}
\pi(X_{\infty})
=\dfrac{1}{2}\left\{\pi(x_{+})+\pi(x_{-})\right\}+2\omega_{1}\dfrac{Y_{\infty}}{X_{\infty}-X_{0}}+a \pi i\,.
\end{equation}
As will seen later,
the exact value of the integer $a$ does not matter in our derivation of the K\"ahler potential for
Atiyah-Hitchin manifold (see \eqref{eqn:fnresult_K} for the expression of the K\"ahler potential).

Substituting \eqref{eqn:omegaomega'_XY} and \eqref{eqn:piinfty_zetaomega'}
 into \eqref{eqn:InGm2}, we obtain
\begin{align}
\mathcal{I}_{1}(\Gamma_{m})
&=
2\beta\omega_{1}
-\beta\dfrac{X_{\infty}-X_{0}}{Y_{\infty}}
\left[
\dfrac{1}{2}\left\{\pi(x_{+})+\pi(x_{-})\right\}+2\omega_{1}\dfrac{Y_{\infty}}{X_{\infty}-X_{0}}+a\pi i
\right] \nonumber \\
&= \dfrac{1}{4\sqrt{z}}
\left\{\pi(x_{+})+\pi(x_{-})
+2a\pi i
\right\}\,.\label{eqn:piinfty_zetaomega''}
\end{align}
Here, in the 
last equality we have used the first equality in
\eqref{eqn:YinfXzeroXinf}.

%
%
\subsection{Calculation of ${\cal I}_2(\Gamma_m)$}
We evaluate \eqref{eqn:InGm} for $n=2$ case, that is,
\begin{equation}\label{eqn:I2Gm1}
\mathcal{I}_{2}(\Gamma_{m})
=\oint_{\Gamma_{m}}\left(\beta\dfrac{X-X_{0}}{X-X_{\infty}}\right)^2\dfrac{dX}{Y}\,.
\end{equation}
We first observe
\begin{equation}\label{eqn:I2Gm'}
\left(\dfrac{X-X_{0}}{X-X_{\infty}}\right)^{2}
=1+\dfrac{2(X_{\infty}-X_{0})}{Y_{\infty}}\cdot\dfrac{Y_{\infty}}{X-X_{\infty}}
+\left(\dfrac{X_{\infty}-X_{0}}{Y_{\infty}}\right)^{2}\left(
\dfrac{Y_{\infty}}{X-X_{\infty}}
\right)^{2}\,,
\end{equation}
so that we will evaluate \eqref{eqn:I2Gm1} by integrating each term in the right-hand side
 of \eqref{eqn:I2Gm'}.
We can get the integrals of the first term and the second term
 by means of \eqref{eqn:omegaomega'_XY} and 
\eqref{eqn:piinfty_zetaomega'}, respectively.
The integral of the second term is calculated as follows:
\begin{align}
\oint_{\Gamma_{m}}
\left\{
\dfrac{2(X_{\infty}-X_{0})}{Y_{\infty}}\cdot
\dfrac{Y_{\infty}}{X-X_{\infty}}\right\}\dfrac{dX}{Y} 
= \dfrac{1}{2\beta\sqrt{z}}\left\{\pi(x_{+})+\pi(x_{-})+2a\pi i\right\}-4\omega_{1}\,.\label{eqn:I2g_2ndt}
\end{align}
In order to perform the integration of the third term,
we observe
\begin{equation}\label{eqn:yxxinfty2}
\left(
\dfrac{Y_{\infty}}{X-X_{\infty}}
\right)^{2}
=2(X-X_{\infty})-\dfrac{12X_{\infty}^{2}-g_{2}}{2Y_{\infty}}\cdot\dfrac{Y_{\infty}}{X-X_{\infty}}-Y\dfrac{d}{dX}\left(\dfrac{Y}{X-X_{\infty}}\right)\,.
\end{equation}
Here we use an integral expression of the quasi-period $\eta_{1}$ of $\zeta$-function
 (see \eqref{eqn:eta1}):
\begin{equation}\label{eqn:quasiintegral}
2\eta_{1}
=-2\int_{e_{3}}^{e_{2}}X\dfrac{dX}{Y}
=-\oint_{\Gamma_{m}}X\dfrac{dX}{Y}\,.
\end{equation}
By using this, calculating the integral of the first term in \eqref{eqn:yxxinfty2} yields:
\begin{align}
\oint_{\Gamma_{m}}
2(X-X_{\infty})\dfrac{dX}{Y}
=-4\eta_{1}-4\omega_{1}\left(
\dfrac{x}{3}-\beta v+2\beta^{2}z
\right)\,.\label{eqn:int_2xxinf}
\end{align}
In the equality in \eqref{eqn:int_2xxinf}, we have used \eqref{eqn:quasiintegral} and
\begin{equation}
X_{\infty}=
\dfrac{x}{3}-\beta v+2\beta^{2}z\,.
\end{equation}
Using the following relation
\begin{equation}\label{eqn:12Xinfty2-g2}
\dfrac{12X_{\infty}^{2}-g_{2}}{4(X_{0}-X_{\infty})}
=\beta v-4\beta^{2}z\,,
\end{equation}
we can calculate the integral of the second term in \eqref{eqn:yxxinfty2}:
\begin{align}
\oint_{\Gamma_{m}}
&\dfrac{12X_{\infty}^{2}-g_{2}}{2Y_{\infty}}\cdot\dfrac{Y_{\infty}}{X-X_{\infty}}\dfrac{dX}{Y}\notag \\
&=-\dfrac{1}{2}\left(
\dfrac{v}{\sqrt{z}}-4\beta\sqrt{z}\right)
\left\{\pi(x_{+})+\pi(x_{-})+2a\pi i\right\}+4\omega_{1}(\beta v-4\beta^{2}z)\,.\label{eqn:int_12xgxx}
\end{align}
The integral of the third term in \eqref{eqn:yxxinfty2} can be evaluated as
\begin{equation}\label{eqn:int_yddx}
\oint_{\Gamma_{m}}
Y\dfrac{d}{dX}\left(\dfrac{Y}{X-X_{\infty}}\right)
\dfrac{dX}{Y}=
\oint_{\Gamma_{m}}
d\left(\dfrac{Y}{X-X_{\infty}}\right)=0\,.
\end{equation}
From \eqref{eqn:int_2xxinf}, \eqref{eqn:int_12xgxx} and \eqref{eqn:int_yddx},
we obtain
\begin{align}
&\oint_{\Gamma_{m}}
\left(\dfrac{Y_{\infty}}{X-X_{\infty}}\right)^{2}\dfrac{dX}{Y}\notag\\
&\phantom{hoge}
=-4\eta_{1}-4\omega_{1}\left(
\dfrac{x}{3}-2\beta^{2}z
\right)
+\dfrac{1}{2}\left(
\dfrac{v}{\sqrt{z}}-4\beta\sqrt{z}
\right)\left\{\pi(x_{+})+\pi(x_{-})+2a\pi i\right\}\label{eqn:int_Yinfxxinfdxy}\,.
\end{align}
Thus, by using \eqref{eqn:I2g_2ndt} and \eqref{eqn:int_Yinfxxinfdxy},
we conclude
\begin{align}
&\mathcal{I}_{2}(\Gamma_{m})\notag\\
&=\beta^{2}\left[
2\omega_{1}+\oint_{\Gamma_{m}}
\left(
\dfrac{2(X_{\infty}-X_{0})}{Y_{\infty}}\cdot
\dfrac{Y_{\infty}}{X-X_{\infty}}\right)\dfrac{dX}{Y}
+\left(\dfrac{X_{\infty}-X_{0}}{Y_{\infty}}\right)^{2}\oint_{\Gamma_{m}}\left(
\dfrac{Y_{\infty}}{X-X_{\infty}}
\right)^{2}\dfrac{dX}{Y}
\right] \notag\\
&=
-\dfrac{1}{z}\left[
\eta_{1}+\omega_{1}\cdot\dfrac{x}{3}-\dfrac{1}{8}\dfrac{v}{\sqrt{z}}\{\pi(x_{+})+\pi(x_{-})+2a\pi i\}
\right]\label{eqn:I2Gm_last}\,.
\end{align}

%
%
\subsection{The function $F$ in terms of elliptic integrals}
We return to our calculation of $F_{1}$.
We will calculate the second term in \eqref{eqn:F1_exp}
 by using \eqref{eqn:piinfty_zetaomega''} and \eqref{eqn:I2Gm_last}.
We get
\begin{align}
v\mathcal{I}_{1}&(\Gamma_{m})-z\mathcal{I}_{2}(\Gamma_{m})
-\dfrac{\pi i}{4}\dfrac{v}{\sqrt{z}}\notag\\
&=\eta_{1}+\omega_{1}\cdot\dfrac{x}{3}+\dfrac{v}{8\sqrt{z}}
\left\{\pi(x_{+})+\pi(x_{-})\right\}+\dfrac{\pi i}{4}(a-1)\cdot\dfrac{v}{\sqrt{z}}\,.
\end{align}
Thanks to \eqref{eqn:ypm}, we find that the integrands of
\begin{equation}
\pi(x_{+})
=-2\int_{e_{3}}^{e_{2}}
\dfrac{y_{+}}{X-x_{+}}\dfrac{dX}{Y},
\quad
\pi(x_{-})
=-2\int_{e_{3}}^{e_{2}}
\dfrac{y_{-}}{X-x_{-}}\dfrac{dX}{Y}\,,
\end{equation}
are pure imaginary and real, respectively, so that we have
\begin{equation}
\pi(x_{+})\in i\mathbb{R},
\quad
\pi(x_{-})\in\mathbb{R}\,.
\end{equation}
This obeys
\begin{align}
\dfrac{v}{\sqrt{z}}
\left\{
\pi(x_{+)}+\pi(x_{-})
\right\}+\mathrm{c.c.}
=2\left\{
iv_{+}\pi(x_{+})+v_{-}\pi(x_{-})
\right\}\,,\label{eqn:vzpp}
\end{align}
where $v_\pm$ are defined in (\ref{eqn:vpvm_def}).
Hence we obtain
\begin{align}
v\mathcal{I}_{1}&(\Gamma_{m})-z\mathcal{I}_{2}(\Gamma_{m})
-\dfrac{\pi i}{4}\dfrac{v}{\sqrt{z}}+\mathrm{c.c.}\notag\\
&=2\eta_{1}+\omega_{1}\cdot\dfrac{2x}{3}
+\dfrac{1}{4}\left\{
iv_{+}\pi(x_{+})+v_{-}\pi(x_{-})
\right\}
-\dfrac{\pi}{2}(a-1)v_{+}\,.\label{eqn:F12nd}
\end{align}
Here, we have used that $\eta_{1}$ is real as noted in (\ref{eqn:eta_expression}) below.
By using \eqref{eqn:F12nd}, $F_{1}$ expressed in \eqref{eqn:F1_exp}
 is rewritten as
\begin{align}
F_{1}
=-8\eta_{1}+8(x_{+}+x_{-})\omega_{1}
-\{iv_{+}\pi(x_{+})+v_{-}\pi(x_{-})\}
+2\pi(a-1)v_{+}\,.
\end{align}
Here, we have used $x=6(x_{+}+x_{-})$.
Thus, the final result for $F=F_{2}+F_{1}$ is
\begin{align}
F=-8\eta_{1}+\left(8\omega_{1}-\dfrac{3}{2h}\right)(x_{+}+x_{-})
-\{iv_{+}\pi(x_{+})+v_-\pi(x_{-})\}+2\pi(a-1)v_{+}
\,.\label{eqn:fresultF}
\end{align} 

%
%
\subsection{Deriving the K\"ahler potential by the generalized Legendre transformation}
The K\"ahler potential $K=K(z,\bar{z},u,\bar{u})$ for the Atiyah-Hitchin manifold
 has the following expression by the generalized Legendre transformation:
\begin{equation}\label{eqn:Kpot_AH}
K(z,\bar{z},u,\bar{u})
=F(z,\bar{z},v,\bar{v},x)-(uv+\bar{u}\bar{v})\,.
\end{equation}
%
with the conditions \eqref{Fv} and \eqref{Ft}, the latter of which, in our setting, reduces to
\begin{align}
\dfrac{\partial F}{\partial x}=0\label{eqn:dfdx}\,.
\end{align}
Then, $K$ satisfies hyperk\"ahler Monge-Amp\`ere equation (cf.~\cite[(4.5)]{Bielawski}):
\begin{equation}\label{eqn:hMAeq}
\det\left(\begin{array}{cc}
K_{z\bar{z}} & K_{z\bar{u}} \\
K_{u\bar{z}} & K_{u\bar{u}}
\end{array}\right)=1\,.
\end{equation}
Let us consider the condition \eqref{Fv}.
By \eqref{eqn:f2_xh} we get
\begin{equation}
\dfrac{\partial F_{2}}{\partial v}
=\dfrac{\partial}{\partial v}\left(-\dfrac{x}{h}\right)=0\,.
\end{equation}
From the third equality in \eqref{eqn:delF1} we find
\begin{align}
\dfrac{\partial F_{1}}{\partial v}
=-\dfrac{1}{2}\cdot\dfrac{1}{\sqrt{z}}\left\{\pi(x_{+})+\pi(x_{-})\right\}
-\dfrac{\pi i}{\sqrt{z}}(a-1)\,,
\end{align}
%
where we have used \eqref{eqn:piinfty_zetaomega''}.
Hence we have
\begin{equation}\label{eqn:u_gLe}
u
=-\dfrac{1}{2}\cdot\dfrac{1}{\sqrt{z}}\left\{\pi(x_{+})+\pi(x_{-})\right\}-\dfrac{(a-1)\pi i}{\sqrt{z}}\,.
\end{equation}
By using this expression we get
\begin{align}
uv+\bar{u}\bar{v}
=-\left\{
iv_{+}\pi(x_{+})+v_{-}\pi(x_{-})
\right\}+2(a-1)\pi v_{+}\label{eqn:uvuvb_exp}\,,
\end{align}
where we have used \eqref{eqn:vzpp}.
In a similar manner, the condition \eqref{eqn:dfdx} is rewritten as
\begin{equation}\label{eqn:dfdx2}
\dfrac{1}{h}=4\omega_{1}\,.
\end{equation}
Indeed, from the last equality in \eqref{eqn:delF1} we have
\begin{equation}\label{eqn:uvubvb2}
\dfrac{\partial F_{1}}{\partial x}
=\mathcal{I}_{0}=4\omega_{1}\,,\quad
\dfrac{\partial F_{2}}{\partial x}
=-\dfrac{1}{h}\,.
\end{equation}
Thus, by substituting \eqref{eqn:fresultF}, \eqref{eqn:uvuvb_exp}
and \eqref{eqn:dfdx2} into \eqref{eqn:Kpot_AH}
we obtain the final expression of the K\"ahler potential
\begin{equation}\label{eqn:fnresult_K}
K=-8\eta_{1}+2(x_{+}+x_{-})\omega_{1}\,.
\end{equation}
This result is consistent with one in \cite{Ionas1}, where differences are coefficients
 and an overall sign. The former stems from the fact that we choose the double contour while the
 latter is a due to a difference of the overall sign of the $F$-function. It should be stressed that this
 K\"ahler potential is real-valued, which is consistent with the definition of the K\"ahler potential.
 This is the consequence of the choice of the double contour. 
 The differences are also reflected to the K\"ahler metric as will be shown below.

%
%
\subsection{Deriving the K\"ahler metric}\label{sec:der_KM}
Let us introduce holomophic coordinates $Z$, $U$ defined by
\begin{equation}
Z=2\sqrt{z}\,,\quad
U=u\sqrt{z}\,.
\end{equation}
We remark that 
this coordinate change preserves hyperk\"ahler Monge-Amp\`ere equation \eqref{eqn:hMAeq}, namely,
\begin{equation}\label{eqn:hMAeq2}
\det\left(\begin{array}{cc}
K_{Z\bar{Z}} & K_{Z\bar{U}} \\
K_{U\bar{Z}} & K_{U\bar{U}}
\end{array}\right)=1\,.
\end{equation}
In this subsection, we will derive the components
$K_{Z\bar{Z}}$,
$K_{U\bar{Z}}$,
$K_{Z\bar{U}}$
and
$K_{U\bar{U}}$
of the metric with respect to the coordinates $(Z,U)$.
Note that in this subsection, we derive the components by differentiating 
 the K\"ahler potential $K$ not using the formula \eqref{eqn:Kzzbar-F}-\eqref{eqn:Kuubar-F},
 where the function $F$ is directly differentiated.
The latter also yields the same results given below though we do not show it in this paper.

We start with evaluating $d\eta_{1}$ and $dx_{\pm}$
 by means of $dZ$, $d\bar{Z}$, $dU$ and $d\bar{U}$.
It follows from \eqref{eqn:u_gLe}
 that $U$ and $\bar{U}$ have the following expressions:
\begin{equation}
U
= 
-\dfrac{1}{2}\left\{\pi(x_{+})+\pi(x_{-})\right\}-\pi i (a-1)
\,,\quad
\bar{U}
= 
-\dfrac{1}{2}\left\{-\pi(x_{+})+\pi(x_{-})\right\}+\pi i(a-1)\,,
\end{equation}
where $a$ is the integer given by \eqref{eqn:piinfty_zetaomega'}.
This obeys
\begin{equation}\label{eqn:dUdUb_dpi}
dU=-\dfrac{1}{2}\{d\pi(x_{+})+d\pi(x_{-})\},\quad
d\bar{U}=-\dfrac{1}{2}\{-d\pi(x_{+})+d\pi(x_{-})\}\,.
\end{equation}
Here $d\pi(x_\pm)$ can be expressed in terms of $dx_\pm$, $dg_2$ and $dg_3$.
Then, $dg_2$ and $dg_3$ can be converted to $d\omega_1$ and $d\eta_1$.
The detailed calculations to obtain $d\pi(x_\pm)$ are summarized in Appendices 
 \ref{sec:domegadeta_eva} and \ref{sec:evaluate_dpi}.
The result is 
\begin{align}
d\pi(x_{\pm})
=\dfrac{4(x_{\pm}\omega_{1}+\eta_{1})}{y_{\pm}}dx_{\pm}
+\dfrac{8(x_{\pm}^{2}-V\eta_{1})}{y_{\pm}}d\omega_{1}-\dfrac{8(x_{\pm}+V\omega_{1})}{y_{\pm}}d\eta_{1}\,, 
\end{align}
where
\begin{equation}\label{eqn:Vdef}
V=\dfrac{-3g_{3}\omega_{1} +2g_{2}\eta_{1}}{12\eta_{1}^{2}-g_{2}\omega_{1}^{2}}\,.
\end{equation}
Using \eqref{eqn:dfdx2}, we have
\begin{align}
d\pi(x_{\pm})=4A_{\pm}dx_{\pm}-8B_{\pm}d\eta_{1}\,,
\end{align}
where we put
\begin{equation}
A_{\pm}=\dfrac{x_{\pm}\omega_{1}+\eta_{1}}{y_{\pm}},\quad
B_{\pm}=\dfrac{x_{\pm}+V\omega_{1}}{y_{\pm}}\,.
\end{equation}
Substituting this into \eqref{eqn:dUdUb_dpi} and making  $dU-d\bar{U}$ and $dU+d\bar{U}$,
 we obtain
\begin{align}
dU-d\bar{U} &= -4A_{+}dx_{+}+8B_{+}d\eta_{1}\,,\\
dU+d\bar{U} &= -4A_{-}dx_{-}+8B_{-}d\eta_{1}\,.
\end{align}
By the definition, we get $|Z|^{2}=x_{+}-x_{-}$.
This yields
\begin{equation}
\bar{Z}dZ+Zd\bar{Z}=dx_{+}-dx_{-}\,.
\end{equation}
By using this, we obtain
\begin{align}
A_{-}(dU-d\bar{U})-A_{+}(dU+d\bar{U})
=-4A_{+}A_{-}(\bar{Z}dZ+Zd\bar{Z})+8(A_{-}B_{+}-A_{+}B_{-})d\eta_{1}\,.
\end{align}
This yields
\begin{equation}
d\eta_{1}
=\dfrac{A_{-}(dU-d\bar{U})-A_{+}(dU+d\bar{U})
+4A_{+}A_{-}(\bar{Z}dZ+Zd\bar{Z})}{8(A_{-}B_{+}-A_{+}B_{-})}\,.
\end{equation}
Furthermore, we have
\begin{align}
(-B_{+}+B_{-})dU-(B_{+}+B_{-})d\bar{U}
=4(A_{-}B_{+}-A_{+}B_{-})dx_{+}-4A_{-}B_{+}(\bar{Z}dZ+Zd\bar{Z})\,,
\end{align}
equivalently,
\begin{equation}
dx_{+}=\dfrac{(-B_{+}+B_{-})dU-(B_{+}+B_{-})d\bar{U}
+4A_{-}B_{+}(\bar{Z}dZ+Zd\bar{Z})}{4(A_{-}B_{+}-A_{+}B_{-})}\,.
\end{equation}
A similar calculation shows
\begin{equation}
dx_{-}=\dfrac{(-B_{+}+B_{-})dU-(B_{+}+B_{-})d\bar{U}
+4A_{+}B_{-}(\bar{Z}dZ+Zd\bar{Z})}{4(A_{-}B_{+}-A_{+}B_{-})}\,.
\end{equation}

Now we are ready to evaluate the components $K_{Z\bar{Z}}$, $K_{U\bar{Z}}$,
 $K_{Z\bar{U}}$ and $K_{U\bar{U}}$.
Introducing the notation
\begin{align}
\mathcal{Q}
&=(\eta_{1}+e_{1}\omega_{1})(\eta_{1}+e_{2}\omega_{1})(\eta_{1}+e_{3}\omega_{1}) \notag \\
&=\eta_{1}^{3}-\dfrac{g_{2}}{4}\omega_{1}^2\eta_{1}+\dfrac{g_{3}}{4}\omega_{1}^{3}\,.
\end{align}
we will have
\begin{align}
K_{Z\bar{Z}}
&=-\dfrac{2}{\mathcal{Q}|Z|^{2}}\mathcal{K}_{4}\,,\label{eqn:KZZb}\\
K_{U\bar{Z}}
&=\dfrac{v_{-}\mathcal{K}_{3+}+iv_{+}\mathcal{K}_{3-}}{2\mathcal{Q}\bar{Z}}\,,\label{eqn:KUZb}\\
\mathcal{K}_{Z\bar{U}}
&=\dfrac{v_{-}\mathcal{K}_{3+}-iv_{+}\mathcal{K}_{3-}}{2\mathcal{Q}Z}\,,\label{eqn:KZUb}\\
\mathcal{K}_{U\bar{U}}
&=-\dfrac{1}{2\mathcal{Q}|Z|^{2}}\mathcal{K}_{2}\,,\label{eqn:KUUb}
\end{align}
where
\begin{align}
\mathcal{K}_{2}
&=\left(\dfrac{g_{2}}{4}-3x_{+}x_{-}\right)\eta_{1}^{2}
-\left\{\dfrac{3g_{3}}{2}+(x_{+}+x_{-})\dfrac{g_{2}}{2}\right\}\omega_{1}\eta_{1}\notag\\
&\phantom{hogehogehogehogehogehogehogehogeho}+\left\{\dfrac{g_{2}^{2}}{16}+3(x_{+}+x_{-})\dfrac{g_{3}}{4}+x_{+}x_{-}\dfrac{g_{2}}{4}\right\}\omega_{1}^{2}\,,\\
\mathcal{K}_{3\pm}
&=\eta_{1}^{3}+3x_{\pm}\omega_{1}\eta_{1}^{2}+\dfrac{g_{2}}{4}\omega_{1}^{2}\eta_{1}-\left(
\dfrac{g_{3}}{2}+x_{\pm}\dfrac{g_{2}}{4}
\right)\omega_{1}^{3}\,,\\
\mathcal{K}_{4}
&=\eta_{1}^{4}+2(x_{+}+x_{-})\omega_{1}\eta_{1}^{3}+\left(\dfrac{g_{2}}{4}+3x_{+}x_{-}\right)
\omega_{1}^{2}\eta_{1}^{2}-\dfrac{g_{3}}{2}\omega_{1}^{3}\eta_{1}\notag\\
&\phantom{hogehogehogehogehogehogehogehogehogehoge}-\left\{
(x_{+}+x_{-})\dfrac{g_{3}}{4}+x_{+}x_{-}\dfrac{g_{2}}{4}
\right\}\omega_{1}^{4}\,.
\end{align}
We start with calculating $K_{Z}$ and $K_{\bar{Z}}$ in terms of
\eqref{eqn:fnresult_K}.
Then, we have
\begin{align}
K_{Z}
&=2\bar{Z}\dfrac{-2A_{+}A_{-}+(A_{-}B_{+}+A_{+}B_{-})\omega_{1}}{A_{-}B_{+}-A_{+}B_{-}} \notag \\
&=-\dfrac{2}{Z}\{2\eta_{1}+(x_{+}+x_{-})\omega_{1}\}\,.
\end{align}
Here
we have used
\begin{align}
A_{-}B_{+}-A_{+}B_{-}&= |Z|^{2}\dfrac{\eta_{1}-V\omega_{1}}{y_{+}y_{-}}\,,\\
-2A_{+}A_{-}+(A_{-}B_{+}+A_{+}B_{-})\omega_{1}
&=-\dfrac{\eta_{1}-V\omega_{1}^{2}}{y_{+}y_{-}}\{2\eta_{1}+(x_{+}+x_{-})\omega_{1}\}\,.
\end{align}
A similar calculation shows
\begin{equation}
K_{\bar{Z}}
=-\dfrac{2}{\bar{Z}}\{2\eta_{1}+(x_{+}+x_{-})\omega_{1}\}\,.
\end{equation}
Hence we obtain
\begin{align}
K_{Z\bar{Z}}
&=-\dfrac{\partial}{\partial Z}K_{\bar{Z}} \notag \\
&=-\dfrac{2(A_{+}A_{-})+2(A_{-}B_{+}+A_{+}B_{-})\omega_{1}}{A_{-}B_{+}-A_{+}B_{-}} \notag \\
&=-\dfrac{2}{|Z|^{2}}\dfrac{12\mathcal{K}_{4}}{(\eta_{1}-V\omega_{1}^{2})(12\eta_{1}^{2}-\omega_{1}^{2}g_{2})} \notag \\
&=-\dfrac{2}{\mathcal{Q}|Z|^{2}}\mathcal{K}_{4}\,,
\end{align}
where, in the last equality, we have used
\begin{equation}
\eta_{1}-V\omega_{1}^{2}=\dfrac{12\mathcal{Q}}{12\eta_{1}^{2}-\omega_{1}^{2}g_{2}}\,.
\end{equation}
A similar calculation shows \eqref{eqn:KUZb} and \eqref{eqn:KZUb}.
Substituting \eqref{eqn:KZZb}--\eqref{eqn:KZUb} into \eqref{eqn:hMAeq2},
 we find that $K_{U\bar{U}}$ is rewritten as
\begin{equation}
K_{U\bar{U}}
=\dfrac{1}{K_{Z\bar{Z}}}(1+K_{Z\bar{U}}K_{U\bar{Z}})
=-\dfrac{1}{2\mathcal{Q}|Z|^{2}}\mathcal{K}_{2}\,.
\end{equation}
The metric obtained has a similar form with the one in \cite{Ionas1}. 
A difference is just a coefficient of the metric. 
Such a difference stems from the choice of contour: We have taken the double contour while
in \cite{Ionas1} the single contour is chosen.
The double contour yields a different coefficient for the first three terms of the $F$-function 
 (\ref{eqn:fresultF})  from one of the single contour case and produces the residue contribution.
The latter does not appear in the single contour case. 
However, the residue contribution does not affect the metric since it cancels through the
 generalized Legendre transformation (\ref{eqn:Kpot_AH}).
One can also understand this when deriving the metric by using (\ref{eqn:Kzzbar-F})-(\ref{eqn:Kuubar-F}) 
 with (\ref{eqn:fresultF}), where the $F$-function is directly differentiated. 
Since the residue is linear in $v$ and  holomorphic in $z$ and vanishes by derivative of $F$ with respect
 to them, it does not contribute to the metric.

%
%
\section{Conclusion}\label{sec:con}
We have restudied the construction of the Atiyah-Hitchin manifold in the generalized Legendre transform approach.
The $F$-functions for the Atiyah-Hitchin manifold are given in \cite{IR} and \cite{Ionas1}, but
 the contours are different, which might yield a discrepancy of the K\"ahler potential and the metric.
We have shown that the choice of the double contour of $F$-function in \cite{IR} actually yields 
 the real K\"ahler potential which is consistent with a definition of a K\"ahler potential.
We have calculated the K\"ahler potential and the K\"ahler metric in terms of holomorphic coordinates 
 by the choice in \cite{IR} for the first time.
They had been only evaluated in the single contour in \cite{Ionas1}.
We have shown all the detailed steps in their derivation, which were missing in \cite{Ionas1, Ionas2}.
In the derivation of the K\"ahler potential, the calculations of the integrations 
 ${\cal I}_n(\Gamma_m)~(n=0,1,2)$ in the $F$-function are necessary.
We have performed them by using the theory of Weierstrass elliptic function.
The necessary formulas related to the Weierstrass elliptic functions
 and the other relations have been explained in a comprehensive way.
In the calculation of the K\"ahler metric, the main result one has to use is the differentiation
 of the Weierstrass elliptic integrals, which has been also given in detail in our paper. 
We have shown that the resultant K\"ahler potential and metric 
 are slightly different from the ones in \cite{Ionas1}.
A difference is a coefficient of the K\"ahler potential and the metric.
This stems from the choice of contour.
Now the K\"ahler potential and the metric have been obtained in terms of the holomorphic
 coordinates and so one of the complex structures is manifest. Such a construction has a potential application
 to fields in geometry and physics, which may be found elsewhere in the future.

\vspace{10mm}
%
%
\noindent {\bf \large Acknowledgements} \\
\noindent The work of M.A. is supported in part by JSPS Grant-in-Aid for Scientific  
Research KAKENHI Grant No. JP21K03565.

%
%
\vspace{1cm}
\noindent {\bf Declaration} \\
\noindent {\bf Conflict of interest} The authors have no relevant financial or non-financial interests to disclose.

\vspace{10mm}
%
%
\appendix
\renewcommand{\theequation}{A.\arabic{equation} }
\setcounter{equation}{0}
\noindent {\Large \bf Appendix}
%
%
\section{Review of Weierstrass $\wp$-function, $\zeta$-function and $\sigma$-functions}\label{sec:Weierstrass}
We first review Weierstrass $\wp$-functions.
Let $\Lambda$ denote the orthogonal lattice in $\mathbb{C}$ defined by
$\Lambda=\mathbb{Z}\cdot 2\omega\oplus \mathbb{Z}\cdot 2\omega'$
where $\omega$ and $\omega^\prime$ are in  \eqref{eqn:omegaKomega'K'}.
The half-periods $\omega$ and $\omega^\prime$ satisfy $\mathrm{Im}(\omega'/\omega)>0$.
The Weierstrass $\wp$-function $\wp(u)=\wp(u, \Lambda)$ is defined by
\begin{equation}
\wp(u)=\dfrac{1}{u^{2}}+\sum_{\lambda\in\Lambda-\{0\}}
\left\{
\dfrac{1}{(u-\lambda)^{2}}-\dfrac{1}{\lambda^{2}}\
\right\}\,,
\quad
u\in\mathbb{C}\,.
\end{equation}
This function is even and has the double periodicity, that is,
\begin{equation}\label{eqn:wp_dp}
\wp(u+2\omega)=\wp(u)\,,\quad
\wp(u+2\omega')=\wp(u)\,,\quad
u\in\mathbb{C}\,.
\end{equation}
We also have the following differential equation:
\begin{equation}
(\wp'(u))^{2}=4\wp(u)^{3}-g_{2}\wp(u)-g_{3}\,.
\end{equation}
We denote by $C^{*}$ the projectivization of $C$, i.e.,
\begin{equation}
C^{*}
=\{[x_{0},x_{1},x_{2}]\in\mathbb{C}P^{2}
\mid
x_{0}x_{2}^{2}=4x_{1}^{3}-g_{2}x_{0}^{2}x_{1}-g_{3}x_{0}^{3}\}\,.
\end{equation}
Thanks to \eqref{eqn:wp_dp}, the $\wp$-function
 induces a function on the torus $\mathbb{C}/\Lambda$, which we write the same symbol $\wp(u)$. 
Then, we obtain a map $\psi:\mathbb{C}/\Lambda\to\mathbb{C}P^{2}$ defined by
\begin{equation}\label{eqn:abel}
\psi(u)=[1, \wp(u),\wp'(u)]=[1, X,Y]\,,\quad
u\in\mathbb{C}/\Lambda\,.
\end{equation}
From Abel's theorem, $\psi$ gives an isomorphism between $\mathbb{C}/\Lambda$
and $C^{*}$.
We call this map the abel map of $C^{*}$.
Then the differential $du$ on $\mathbb{C}/\Lambda$ is mapped to the abel form $\varpi=dX/Y$.
The inverse map of $\psi$ is given by
\begin{equation}\label{eqn:inverse}
u=\psi^{-1}(p)=\int_{\infty}^{p}\dfrac{dX}{Y},
\quad
p\in C\,,
\end{equation}
which is defined modulo $\Lambda$.
Here, $\infty$ corresponds to the point $[0,0,1]$ of $C^{*}$.
We set $\omega_{1}=\omega$, $\omega_{2}=\omega+\omega'$ and $\omega_{3}=\omega'$ as in \eqref{eqn:omegadef}.
It is verified that $e_{i}=\wp(\omega_{i})$ holds for $i=1,2,3$.
Indeed, from
\begin{equation}\label{eqn:omega1integral}
\omega_{1}=\int_{\infty}^{e_{1}}\dfrac{dX}{Y}\,,\quad
\omega_{3}=\int_{\infty}^{e_{3}}\dfrac{dX}{Y}\,\quad
(\text{mod $\Lambda$})\,,
\end{equation}
we get $e_{1}=\wp(\omega_{1})$, $e_{3}=\wp(\omega_{3})$.
We also get $e_{2}=\wp(\omega_{2})$ by means of the following
formula (cf.~\cite[p.~332]{ht2}):
\begin{equation}\label{eqn:wpupw}
\det\left(
\begin{array}{ccc}
1 & \wp(u) & \wp'(u)\\
1 & \wp(v) & \wp'(v)\\
1 & \wp(w) & \wp'(w)\\
\end{array}
\right)=0\,,\quad
u+v+w=0\,.
\end{equation}
Indeed, by applying $(u,v,w)=(\omega_{1},-\omega_{2},\omega_{3})$ 
 to this formula we have
\begin{equation}
\det\left(
\begin{array}{ccc}
1 & e_{1} & 0\\
1 & \wp(\omega_{2}) & -\wp'(\omega_{2})\\
1 & e_{3} & 0\\
\end{array}\right)=0\,,
\end{equation}
that is, $\wp'(\omega_{2})=0$.
Here, we have used that $\wp'(u)$ is odd.
This means that
$\wp(\omega_{2})$ is a root of $4X^{3}-g_{2}X-g_{3}=0$.
In addition, $\wp(\omega_{2})$ is different from $e_{1}$ and $e_{3}$.
Thus we have $e_{2}=\wp(\omega_{2})$.

Next we review Weierstrass $\zeta$-function.
The $\zeta$-function is defined by the following formula (cf.~\cite[p.~35]{Ak}):
\begin{equation}\label{eqn:zeta_af}
\zeta(u)
=-\int_{0}^{u}\wp(u)du\,,
\end{equation}
so that $\zeta'(u)=-\wp(u)$ holds.
Then the $\zeta$-function is odd.
It is also verified that this function has the following
 double periodicity (cf.~\cite[p.~35]{Ak}):
If we put $\eta_{1}=\zeta(\omega_{1})$ and $\eta_{3}=\zeta(\omega_{3})$, then
\begin{equation}\label{eqn:eta_dperiod}
\zeta(u+2\omega_{1})=\zeta(u)+2\eta_{1}\,,\quad
\zeta(u+2\omega_{3})=\zeta(u)+2\eta_{3}\,,\quad
u\in\mathbb{C}\,.
\end{equation}
%
It follows from $\omega_2=\omega_1+\omega_3$ that
$\eta_{2}=\zeta(\omega_{2})$ satisfies the following relation:
\begin{equation}\label{eqn:eta2}
\eta_{2}=\eta_{1}+\eta_{3}\,.
\end{equation}
%
%
%
%
%
Here, $\eta_i (i=1,2,3)$ is called a quasi-half period, whose integral expression is obtained from
 \eqref{eqn:zeta_af} with the abel map \eqref{eqn:abel} and the inverse map \eqref{eqn:inverse}. 
  For instance, $\eta_1$ is expressed as
\begin{eqnarray}\label{eqn:eta1}
 \eta_1=\zeta(\omega_1)=-\int_{\omega_3}^{\omega_2}\wp(u)du=-\int_{e_3}^{e_2}X{dX \over Y}\,.
\end{eqnarray}
%
%
It can be verified that $\eta_{1}$ 
 (the proof is given in Appendix B in \cite{arxiv:9706145}):
\begin{equation}\label{eqn:eta_expression}
\eta_{1}=\dfrac{-1}{\sqrt{e_{1}-e_{3}}}\{e_{1}K(k)-(e_{1}-e_{3})E(k)\}\,,
\end{equation}
where $E(k)$ denotes the complete elliptic integral of the second kind with modulus $k$.
From this expression, we find that $\eta_{1}$ is in $\mathbb{R}$.

Lastly we review Weierstrass $\sigma$-function, $\sigma(u)=\sigma(u,\Lambda)$,
This is defined by the following equality (cf.~\cite[p.~37]{Ak}):
\begin{equation}
\log\dfrac{\sigma(u)}{u}
=\int_{0}^{u}\left\{
\zeta(u)-\dfrac{1}{u}
\right\}du\,,
\end{equation}
so that
\begin{equation}
\zeta(u) =\dfrac{d}{du}\log \sigma(u)\,. \label{eqn:sigma}
\end{equation}
It is shown that the $\sigma$-function is odd.
Weierstrass $\sigma$-function satisfies the monodromy property, which is derived as follows:
By the double periodicity \eqref{eqn:eta_dperiod} for the $\zeta$-function, we have
\begin{equation}
\dfrac{\sigma'(u+2\omega_{j})}{\sigma(u+2\omega_{j})}
=\zeta(u+2\omega_{j})=\zeta(u)+2\eta_{j}=\dfrac{\sigma'(u)}{\sigma(u)}+2\eta_{j},\quad
j=1,3.
\end{equation}
This implies that
\begin{equation}\label{eqn:monod_j1,3}
\sigma(u+2\omega_{j})=-e^{2\eta_{j}(u+\omega_{j})}\sigma(u),\quad
j=1,3.
\end{equation}
Following to \cite[(5), p.~36]{Ak}
it is shown that
\begin{equation}
\eta_{1}\omega_{3}-\eta_{3}\omega_{1}=\dfrac{\pi i}{2}\,.
\end{equation}
By using this, \eqref{eqn:monod_j1,3} yields the same relation for $j=2$.

%
%
\renewcommand{\theequation}{B.\arabic{equation} }
\setcounter{equation}{0}
\section{Proofs of relations}\label{sec:pr}
%
%
\subsection{Proof of \eqref{eqn:InbarIn}}\label{sec:InbarIn}
We start to define the function as
\begin{equation}
f_{n}(\zeta)=\zeta^{n}\dfrac{1}{2\zeta\sqrt{\eta^{(4)}(\zeta)}}.
\end{equation}
Then we have
\begin{equation}\label{eq:Im_cpxconj}
\overline{\mathcal{I}_{n}}
=\overline{\oint_{\Gamma}f_{n}(\zeta)d\zeta}
=\overline{\oint_{\Gamma_{m}}f_{n}(\zeta)d\zeta}
+
\overline{\oint_{\Gamma_{m}'}f_{n}(\zeta)d\zeta}.
\end{equation}
Set $g_n(\bar{\zeta})=\overline{f_n(\zeta)}$ and $\zeta'=-1/\zeta$.
The first term of  the right-hand side of \eqref{eq:Im_cpxconj} is rewritten as
\begin{align}
\overline{\oint_{\Gamma_{m}}f_{n}(\zeta)d\zeta}
=2\int_{\bar{\alpha}}^{-1/\beta}g_{n}(\bar{\zeta})d\bar{\zeta} 
=2\int_{-1/\bar{\alpha}}^{\beta}g_{n}\left(-\dfrac{1}{\bar{\zeta'}}\right)
\dfrac{d\bar{\zeta'}}{\bar{\zeta'}^{2}}.\label{eqn:intG1_cpxconj}
\end{align}
By using the reality condition (\ref{real}),
\begin{equation}
\overline{
\eta^{(4)}(-1/\zeta')}
=\eta^{(4)}(\bar{\zeta'})\,,
\end{equation}
we have
\begin{align}
g_{n}\left(-\dfrac{1}{\bar{\zeta'}}\right)
=(-1)^{-n+1}\bar{\zeta'}^{-n+2}\dfrac{1}{2\bar{\zeta'}\sqrt{\eta^{(4)}(\bar{\zeta'})}}
=(-1)^{-n+1}f_{-n+2}(\bar{\zeta'}).\label{eqn:gm_f-m+2}
\end{align}
Substituting \eqref{eqn:gm_f-m+2} into \eqref{eqn:intG1_cpxconj}, we obtain
\begin{align}
\overline{\oint_{\Gamma_{m}}f_{n}(\zeta)d\zeta}
=2\int_{-1/\bar{\alpha}}^{\beta}(-1)^{-n+1}f_{-n+2}(\bar{\zeta'})\dfrac{d\bar{\zeta'}}{\bar{\zeta'}^{2}}
=(-1)^{n}\oint_{\Gamma_{m}'}f_{-n}(\zeta)d\zeta\,. \label{eqn:intG1_cpxconj'}
\end{align}
In a similar manner, we find
\begin{equation}\label{eqn:intG1'_cpxconj}
\overline{\oint_{\Gamma_{m}'}f_{n}(\zeta)d\zeta}
=(-1)^{n}\oint_{\Gamma_{m}}f_{-n}(\zeta)d\zeta.
\end{equation}
Hence, by substituting
\eqref{eqn:intG1_cpxconj'}
and
\eqref{eqn:intG1'_cpxconj}
into
\eqref{eq:Im_cpxconj} we have \eqref{eqn:InbarIn}.
This completes the proof.
\color{black}

%
%
\subsection{Proof of \eqref{eqn:YinfXzeroXinf}}\label{sec:Xinfty-X0Yinfty}
By using \eqref{eqn:X0_def} and \eqref{eqn:Xinfty_def}, we have
\begin{align}
\left(
\dfrac{Y_{\infty}}{X_{0}-X_{\infty}}
\right)^{2}
=\dfrac{4X_{\infty}^{3}-g_{2}X_{\infty}-g_{3}}{(X_{0}-X_{\infty})^{2}} 
= 4\beta^{2}\dfrac{\rho\bar{\alpha}\bar{\beta}}{(1+|\alpha|^{2})(1+|\beta|^{2})}
=4\beta^{2}z\,.
\end{align}
Here, in the last equality we have used \eqref{eqn:z}.
By the choice of the branch cut of the square root
$\sqrt{4X_{\infty}^{3}-g_{2}X_{\infty}-g_{3}}$, we have
\begin{equation}
\dfrac{Y_{\infty}}{X_{0}-X_{\infty}}=2\beta\sqrt{z}\,,
\end{equation}
from which the first equality in \eqref{eqn:YinfXzeroXinf} holds.
A similar argument shows the second equality in \eqref{eqn:YinfXzeroXinf}.

%
%
\subsection{Proof of \eqref{eqn:xpm_escpt}}\label{sec:xpm_escpt}
We start with recalling the following formula:
\begin{equation}
\wp(u+v)
=-(\wp(u)+\wp(v))
+\dfrac{1}{4}\left(
\dfrac{\wp'(u)-\wp'(v)}{\wp(u)-\wp(v)}
\right)^{2}\,,\quad
u,v\in\mathbb{C}\,.
\end{equation}
Substituting $u=u_{\infty}$ and $v=\pm u_{0}$ into this formula, we obtain
\begin{align}\label{eqn:wpxp}
\wp(u_{\infty}^{\pm})
&=-(\wp(u_{\infty})+\wp(u_{0}))
+\dfrac{1}{4}\left(
\dfrac{\wp'(u_{\infty})\mp\wp'(u_{0})}{\wp(u_{\infty})- \wp( u_{0})}
\right)^{2} \notag \\
&=-(X_{\infty}+X_{0})
+\dfrac{1}{4}\left(
\dfrac{Y_{\infty}}{X_{0}-X_{\infty}}
\pm \dfrac{Y_{0}}{X_{\infty}-X_{0}}
\right)^{2}\,.
\end{align}
By using \eqref{eqn:X0_def} and \eqref{eqn:Xinfty_def} it is verified that
the first term in \eqref{eqn:wpxp} is rewritten as
\begin{equation}\label{eqn:1stwpwx}
-(X_{\infty}+X_{0})
= \dfrac{x}{3}-\left(
\beta^{2}z+
\dfrac{\bar{z}}{\beta^{2}}
\right)\,.
\end{equation}
From \eqref{eqn:YinfXzeroXinf}
the second term in \eqref{eqn:wpxp}
is rewritten as
\begin{equation}\label{eqn:2ndwpwx}
\dfrac{1}{4}\left(
\dfrac{Y_{\infty}}{X_{0}-X_{\infty}}
\pm \dfrac{Y_{0}}{X_{\infty}-X_{0}}
\right)^{2}=
\beta^{2}z+\dfrac{\bar{z}}{\beta^{2}}\pm 2|z|\,.
\end{equation}
Combined \eqref{eqn:1stwpwx} and \eqref{eqn:2ndwpwx}, we have
\begin{equation}
x_{\pm}=\wp(u^\pm_\infty)=\dfrac{x\pm 6|z|}{3}\,,
\end{equation}
from which \eqref{eqn:xpm_escpt} holds.

%
%
\renewcommand{\theequation}{C.\arabic{equation} }
\setcounter{equation}{0}
\section{Differential formulas in Weierstrass normal form}\label{sec:dfw}
This appendix provides some formulas, which are used to calculate the metric
of Atiyah-Hitchin manifold in Subsection \ref{sec:der_KM}.
In our derivation of the formulas, we make use of differential formulas
for complete elliptic integrals (see \cite{Ak, BF} for basic references).

%
%
\subsection{Evaluations of $d\omega$ and $d\eta$}\label{sec:domegadeta_eva}
The purpose of this subsection is to evaluate the differentials for $\omega$ and $\eta$.
We start with the curve of the Weierstrass normal form, $C$, in \eqref{eqn:Wcurve}.
The half-period and the quasi-period of $C$ are written by means of integral forms, i.e.,
\begin{align}
\omega
&=\omega(g_{2},g_{3})=\int_{e_{3}}^{e_{2}}\dfrac{dX}{Y}\,,\\
\eta
&=\eta(g_{2},g_{3})=-\int_{e_{3}}^{e_{2}}X\dfrac{dX}{Y}\,.
\end{align}
We note that $\omega$ and $\eta$ are equal to $\omega_{1}$ in \eqref{eqn:omega1integral} 
 above and $\eta_{1}$ in \eqref{eqn:eta_dperiod}, respectively.

%
%
%
%
%
%
%
%
%

We first derive the Jacobian in
\begin{equation}\label{egn:dgdgdrdk}
\left(\begin{array}{c}
dg_{2} \\
dg_{3}
\end{array}\right)=
J\left(\begin{array}{c}
d\rho\\
dk^{2}
\end{array}
\right)\,.
\end{equation}
This can be easily calculated by using \eqref{eqn:g2g3_def}:
\begin{equation}
J=
\dfrac{D(g_{2},g_{3})}{D(\rho,k^{2})}
=\left(\begin{array}{cc}
\dfrac{8\rho}{3}(1-k^{2}+k^{4}) & \dfrac{4\rho^{2}}{3}(-1+2k^{2}) \\
\dfrac{4\rho^{2}}{9}(k^{2}-2)(2k^{2}-1)(k^{2}+1) & \dfrac{4\rho^{3}}{9}(-1-2k^{2}+2k^{4})
\end{array}\right)\,.
\end{equation}
With the use of \eqref{eqn:Deltakrho_apndx} we obtain
\begin{equation}
\det(J)=-\dfrac{\Delta}{3k^{2}k'^{2}\rho^{2}}\,.
\end{equation}
The inverse matrix $J^{-1}$
is given by
\begin{equation}\label{eqn:Jinv_apdx}
J^{-1}
=
-\dfrac{3k^{2}k'^{2}\rho^{2}}{\Delta}
\left(\begin{array}{cc}
\dfrac{4\rho^{3}}{9}(-1-2k^{2}+2k^{4}) & -\dfrac{4\rho^{2}}{3}(-1+2k^{2}) \\
-\dfrac{4\rho^{2}}{9}(k^{2}-2)(2k^{2}-1)(k^{2}+1) & \dfrac{8\rho}{3}(1-k^{2}+k^{4})
\end{array}\right)\,.
\end{equation}

Next, we recall the differentiation formulas for the complete elliptic integrals
 of the first kind $K(k)$ and the second kind $E(k)$ (cf.~\cite{BF}):
%
\begin{equation}\label{eqn:dKdEformula}
\dfrac{dK(k)}{dk^{2}}=\dfrac{E(k)-k'^{2}K(k)}{2k^{2}k'^{2}}\,,\quad
\dfrac{dE(k)}{dk^{2}}=\dfrac{E(k)-K(k)}{2k^{2}}\,.
\end{equation}
%

We are ready to derive our differentiation formulas for $\omega$ and $\eta$.
We first evaluate $d\omega$.
We observe (see Subsection \ref{sec:I0})
\begin{align}
\omega
=\dfrac{1}{\sqrt{\rho}}K(k)\label{eqn:omegaK_apdx}\,.
\end{align}
%
By differentiating the both sides of \eqref{eqn:omegaK_apdx}, we get
\begin{align}\label{eqn:domega_drdk}
d\omega
= -\dfrac{1}{2\rho\sqrt{\rho}}K(k)d\rho+
\dfrac{1}{2\sqrt{\rho}}\dfrac{E(k)-k'^{2}K(k)}{k^{2}k'^{2}}dk^{2}\,,
\end{align}
%
where we have used the first equality in \eqref{eqn:dKdEformula}.
Combined \eqref{eqn:Jinv_apdx} and \eqref{eqn:domega_drdk}
 we can obtain an explicit description of the partial derivatives 
 $\partial \omega/\partial g_{2}$ and $\partial \omega/\partial g_{3}$.
Indeed, we have
\begin{equation}
\dfrac{\partial \omega}{\partial g_{2}}
=\dfrac{\partial \omega}{\partial \rho}\dfrac{\partial \rho}{\partial g_{2}}
+\dfrac{\partial \omega}{\partial k^{2}}\dfrac{\partial k^{2}}{\partial g_{2}}
=\dfrac{-g_{2}^{2}\omega+18g_{3}\eta}{4\Delta}\,.
\end{equation}
By a similar calculation, we get
\begin{equation}
\dfrac{\partial \omega}{\partial g_{3}}
=\dfrac{3(3g_{3}\omega-2g_{2}\eta)}{2\Delta}\,,
\end{equation}
so that \eqref{eqn:domega_drdk} is rewritten as
\begin{equation}\label{eqn:domegadg2dg3}
d\omega
=\dfrac{-g_{2}^{2}\omega+18g_{3}\eta}{4\Delta}dg_{2}
+\dfrac{3(3g_{3}\omega-2g_{2}\eta)}{2\Delta}dg_{3}\,.
\end{equation}

Next, let us evaluate $d\eta$.
From \eqref{eqn:eta_expression}, we have
\begin{align}\label{eqn:etaEK_apdx}
\eta
=\dfrac{\sqrt{\rho}}{3}\left\{
(k^{2}-2)K(k)+3E(k)\right\}\,.
\end{align}
%
whose differentiation yields the following expression:
\begin{equation}\label{eqn:detadg2dg3}
d\eta=\dfrac{-g_{2}(3g_{3}\omega-2g_{2}\eta)}{8\Delta}dg_{2}
-\dfrac{-g_{2}^{2}\omega+18g_{3}\eta}{4\Delta}dg_{3}\,.
\end{equation}


\subsection{Evaluation of $d\pi$}\label{sec:evaluate_dpi}

We introduce the following function $\pi=\pi(x,g_{2},g_{3})$ defined by
\begin{equation}
\pi = \pi(x,g_{2},g_{3})
=-\oint_{\Gamma_{m}}\dfrac{y}{X-x}\dfrac{dX}{Y}\,,
\end{equation}
where $y^{2}=4x^{3}-g_{2}x-g_{3}$ holds.
This integral coincides with that in \eqref{eqn:piGamma}.
In order to derive the differential of $\pi$, we need to calculate $\partial \pi/\partial x$, 
 $\partial \pi/\partial g_2$ and $\partial \pi/\partial g_3$.
To this end, we make use of the results in \cite{arxiv:9706145}. 
In this paper, the following integrals was calculated:
\begin{align}
I_{1}^{(1)}&=I^{(1)}_{1}(g_{2},g_{3})=\oint_{\Gamma_{m}}\dfrac{dX}{Y}\,,\\
I_{2}^{(1)}&=I_{2}^{(1)}(g_{2},g_{3})=\oint_{\Gamma_{m}}X\dfrac{dX}{Y}\,,\\
I_{3}^{(1)}&=I_{3}^{(1)}(x)=I_{3}^{(1)}(x,g_{2},g_{3})=\oint_{\Gamma_{m}}\frac{1}{X-x}\dfrac{dX}{Y}\,,
\end{align}
where $\Gamma_{m}$ is an integration contour encircling $e_{3}$ and $e_{2}$ of $X$-plane.
Then our $\omega(=\omega_1)$, $\eta(=\eta_1)$ and $\pi$ are related with the above expression as
%
\begin{equation}
\omega=\dfrac{1}{2}I^{(1)}_{1}\,\quad
\eta=-\dfrac{1}{2}I^{(1)}_{2}\,\quad
\pi= -y I^{(1)}_{3}
\end{equation}
From \cite[(B.16)]{arxiv:9706145}
 the relation between $I_1^{(1)}$, $I_2^{(1)}$ and $I_3^{(1)}$ is given by
\begin{equation}\label{eqn:I3_delx}
y^{2}\dfrac{\partial I^{(1)}_{3}}{\partial x}
=-2x I^{(1)}_{1}+2I^{(1)}_{2}-\dfrac{1}{2}(12x^{2}-g_{2})I^{(1)}_{3}\,.
\end{equation}
From this we have $\partial \pi/\partial x$ as
\begin{align}
\dfrac{\partial \pi}{\partial x}
=\dfrac{4x\omega+4\eta}{y}\label{eqn:dpidx}\,.
\end{align}
We next evaluate $\partial\pi/\partial g_{2}$.
We observe
\begin{align}
\dfrac{\partial I_{3}^{(1)}}{\partial g_{2}}
=\dfrac{1}{8}\oint_{\Gamma_{m}}\dfrac{X}{(X-x)(X-e_{1})(X-e_{2})(X-e_{3})}\dfrac{dX}{Y}\,.
\end{align}
If we put
\begin{align}
a&=\dfrac{x}{(x-e_{1})(x-e_{2})(x-e_{3})}=\dfrac{4x}{y^{2}}\,,\\
b&=\dfrac{e_{1}}{(x-e_{1})(e_{1}-e_{2})(e_{3}-e_{1})}\,,\\
c&=\dfrac{e_{2}}{(x-e_{2})(e_{1}-e_{2})(e_{2}-e_{3})}\,,\\
d&=\dfrac{e_{3}}{(x-e_{3})(e_{2}-e_{3})(e_{3}-e_{1})}\,,
\end{align}
then we get
\begin{equation}
\dfrac{X}{(X-x)(X-e_{1})(X-e_{2})(X-e_{3})}
=\dfrac{a}{X-x}+\dfrac{b}{X-e_{1}}+\dfrac{c}{X-e_{2}}+\dfrac{d}{X-e_{3}}\,,
\end{equation}
from which we obtain
\begin{equation}
\dfrac{\partial I_{3}^{(1)}}{\partial g_{2}}
=\dfrac{1}{8}\left\{
aI_{3}^{(1)}(x)+bI_{3}^{(1)}(e_{1})+cI_{3}^{(1)}(e_{2})+dI_{3}^{(1)}(e_{3})
\right\}.
\end{equation}
We remark that the integrals $I_{3}^{(1)}(e_{1})$, $I_{3}^{(1)}(e_{2})$ and $I_{3}^{(1)}(e_{3})$
 have the following expression (cf.~\cite[(B.11--13)]{arxiv:9706145}):
\begin{align}
I_{3}^{(1)}(e_{1})&=\dfrac{-2}{(e_{1}-e_{3})^{3/2}}\dfrac{E(k)}{1-k^{2}}\,,\\
I_{3}^{(1)}(e_{2})&=\dfrac{2}{(e_{1}-e_{3})^{3/2}}\dfrac{1}{k^{2}}\left(\dfrac{E(k)}{1-k^{2}}-K(k)\right)\,,\\
I_{3}^{(1)}(e_{3})&=\dfrac{2}{(e_{1}-e_{3})^{3/2}}\dfrac{1}{k^{2}}\left(K(k)-E(k)\right)\,.
\end{align}
Thus we obtain
\begin{align}
\dfrac{\partial \pi}{\partial g_{2}}
&= -\dfrac{\partial y}{\partial g_{2}}I_{3}^{(1)}
-y\dfrac{\partial I_{3}^{(1)}}{\partial g_{2}} \notag \\
&=-\dfrac{1}{4y(e_{1}-e_{3})^{3/2}}\cdot y^{2}\left\{
\dfrac{-c+d}{k^{2}}K(k)
+\left(-\dfrac{b}{1-k^{2}}+\dfrac{c}{k^{2}}\dfrac{1}{1-k^{2}}-\dfrac{d}{k^{2}}\right)E(k)
\right\}\label{eqn:dpig2}\,.
\end{align}
Here, we have used
\begin{equation}
\dfrac{x}{2y}-\dfrac{ay}{8}
=\dfrac{x}{2y}-\dfrac{4x}{y^{2}}\dfrac{y}{8}=0\,.
\end{equation}
Furthermore, a direct calculation shows
\begin{align}
&y^{2}\left\{
\dfrac{-c+d}{k^{2}}K(k)
+\left(-\dfrac{b}{1-k^{2}}+\dfrac{c}{k^{2}}\dfrac{1}{1-k^{2}}-\dfrac{d}{k^{2}}\right)E(k)
\right\}\notag\\
&\phantom{hogehoge}= -\dfrac{1}{4k^{4}k'^{4}\rho^{4}\sqrt{\rho}}\left\{
(g_{2}x+3g_{3})(3g_{3}\omega-2\eta g_{2})+2(-g_{2}^{2}\omega+18g_{3}\eta)x^{2}
\right\}\,.
\end{align}
Hence we obtain
\begin{equation}
\dfrac{\partial \pi}{\partial g_{2}}
=\dfrac{(g_{2}x+3g_{3})(3g_{3}\omega-2\eta g_{2})
+2(-g_{2}^{2}\omega+18g_{3}\eta)x^{2}
}{y\Delta}\label{eqn:dpidg2}\,.
\end{equation}
In a similar manner, $\partial \pi/\partial g_{3}$ is obtained as
%
%
%
%
%
%
%
%
%
\begin{align}\label{eqn:dpidg3}
\dfrac{\partial \pi}{\partial g_{3}}
&=\dfrac{2}{y\Delta}\left\{
(6x^{2}-g_{2})(3g_{3}\omega-2g_{2}\eta)
+(-g_{2}^{2}\omega+18g_{3}\eta)x
\right\}\,.
\end{align}
It follows from \eqref{eqn:dpidx}, \eqref{eqn:dpidg2} and \eqref{eqn:dpidg3} that the 
differential of $\pi$ has the following expression:
\begin{align}
d\pi
&= \dfrac{4x\omega+4\eta}{y}dx
+\dfrac{(g_{2}x+3g_{3})(3g_{3}\omega-2\eta g_{2})
+2(-g_{2}^{2}\omega+18g_{3}\eta)x^{2}
}{y\Delta}dg_{2} \notag\\
&\phantom{hogehogehoge}+\dfrac{2\left\{
(6x^{2}-g_{2})(3g_{3}\omega-2g_{2}\eta)
+(-g_{2}^{2}\omega+18g_{3}\eta)x
\right\}}{y\Delta}dg_{3}\,.
\end{align}
It is convenient to rewrite $dg_{2}$ and $dg_{3}$ by means of $d\omega$ and $d\eta$.
Calculating the Jacobian as
%
\begin{equation}
J'
=\dfrac{D(\omega,\eta)}{D(g_{2},g_{3})}
=\left(
\begin{array}{cc}
\dfrac{-g_{2}^{2}\omega+18g_{3}\eta}{4\Delta} & \dfrac{3(3g_{3}\omega-2g_{2}\eta)}{2\Delta} \\
\dfrac{-g_{2}(3g_{3}\omega-2g_{2}\eta)}{8\Delta} & -\dfrac{-g_{2}^{2}\omega+18g_{3}\eta}{4\Delta}
\end{array}
\right)\,,
\end{equation}
%
we obtain
\begin{equation}
\det(J')=
\dfrac{12\eta^{2}-g_{2}\omega^{2}}{16\Delta}\,.
\end{equation}
In what follows, we assume that $J'$ is non-singular, equivalently,
$12\eta^{2}-\omega^{2}g_{2}\neq 0$.
The inverse matrix $J'^{-1}$ is given by
\begin{equation}
J'^{-1}
=\dfrac{1}{12\eta^{2}-g_{2}\omega^{2}}
\left(
\begin{array}{cc}
4(g_{2}^{2}\omega-18g_{3}\eta) & -24(3g_{3}\omega-2g_{2}\eta) \\
2g_{2}(3g_{3}\omega-2g_{2}\eta) & 
4(-g_{2}^{2}\omega+18g_{3}\eta)
\end{array}
\right)\,.
\end{equation}
From the above argument we obtain the following expression of $d\pi$:
\begin{align}
d\pi
&=\dfrac{\partial \pi}{\partial x}dx
+\left(\dfrac{\partial \pi}{\partial g_{2}}\dfrac{\partial g_{2}}{\partial \omega}+\dfrac{\partial \pi}{\partial g_{3}}\dfrac{\partial g_{3}}{\partial \omega}\right)d\omega
+\left(\dfrac{\partial \pi}{\partial g_{2}}\dfrac{\partial g_{2}}{\partial \eta}+\dfrac{\partial \pi}{\partial g_{3}}\dfrac{\partial g_{3}}{\partial \eta}\right)d\eta \notag \\
&=\dfrac{4x\omega+4\eta}{y}dx
+\dfrac{8(x^{2}-V\eta)}{y}d\omega
-\dfrac{8(x+V\omega)}{y}d\eta\,,\label{eqn:dpidxdomegadeta}
\end{align}
where $V$ is given in \eqref{eqn:Vdef}.

\end{document}